\documentclass[11pt,authoryear]{myarticle}   

\usepackage{setspace}
\usepackage[margin=3cm]{geometry}

\usepackage{array}
\usepackage{booktabs}
\usepackage{lineno,hyperref}
\modulolinenumbers[5]
\usepackage{mathtools}

\usepackage{graphicx}
\usepackage{caption}
\usepackage{xcolor}
\usepackage{subcaption}
\usepackage{natbib}

\usepackage{algorithm}
\usepackage{algcompatible}

\usepackage{amssymb}
\usepackage{bbm}
\usepackage{amsmath}
\usepackage{bm}

\newcommand{\Prt}[1]{\left({#1}\right)}

\newcommand{\Brc}[1]{{\left\{{#1}\right\}}}

\newcommand{\Pmeas}{{\mathbb{P}}}
\newcommand{\Qmeas}{{\mathbb{Q}}}

\newcommand{\Fc}{\mathcal{F}}

\newcommand{\beql}[1]{\begin{equation}\label{#1}}
\newcommand{\eeq}{\end{equation}}
\newcommand{\beqr}[1]
{\begin{equation}\label{#1}\begin{aligned}}
\newcommand{\eeqr}
{\end{aligned}\end{equation}}
\newcommand{\beqc}[1]
{\begin{equation}\label{#1}\begin{cases}}
\newcommand{\eeqc}
{\end{cases}\end{equation}}
\newcommand{\bry}[1]
{\begin{array}{#1}}
\newcommand{\ery}{\end{array}}

\newcommand{\Sh}{\hat{S}}

\newcommand{\bpi}{\bm{\pi}}
\newcommand{\bth}{\bm{\theta}}

\newcommand{\Ff}{\mathbb{F}}

\newcommand{\hf}{\frac{1}{2}}
\newcommand{\bW}{\bm{W}}

\newcommand{\bS}{\bm{S}}
\newcommand{\bX}{\bm{X}}
\newcommand{\eqbyl}{\overset{\mathcal{L}}{=}}

\algnewcommand\INPUT{\item[\textbf{Input:}]}%
\algnewcommand\OUTPUT{\item[\textbf{Output:}]}%

\newcommand{\pt}{t_{n-1}}
\newcommand{\tn}{t_n}
\newcommand{\tp}{t_n^+}

\begin{document}

\begin{frontmatter}

\title{Fair Pricing of Variable Annuities with Guarantees under the Benchmark Approach}

\author[utsm,csiro]{Jin Sun}
\ead{jin.sun@uts.edu.au}

\author[melbaddr]{Kevin Fergusson\corref{corrauth}}
\cortext[corrauth]{Corresponding author}
\ead{kevin.fergusson@unimelb.edu.au}

\author[utsm,utsb]{Eckhard Platen}
\ead{eckhard.platen@uts.edu.au}

\author[mqaddr1]{Pavel V. Shevchenko}
\ead{pavel.shevchenko@mq.edu.au}

\address[utsm]{Faculty of Science, University of Technology Sydney}
\address[melbaddr]{Centre for Actuarial Studies, Department of Economics, University of Melbourne, Victoria, Australia}
\address[csiro]{Data 61, CSIRO Docklands}
\address[mqaddr1]{Department of Actuarial Studies and Business Analytics, Macquarie University, Australia}
\address[utsb]{UTS Business School, University of Technology Sydney}

\begin{abstract}
In this paper we consider the pricing of variable annuities (VAs) with guaranteed minimum withdrawal benefits. We consider two pricing approaches, the classical risk-neutral approach and the benchmark approach, and we examine the associated static and optimal behaviors of both the investor and insurer. The first model considered is the so-called minimal market model, where pricing is achieved using the benchmark approach. The benchmark approach was introduced by Platen in 2001 and has received wide acceptance in the finance community. Under this approach, valuing an asset involves determining the minimum-valued replicating portfolio, with reference to the growth optimal portfolio under the real-world probability measure, and it both subsumes classical risk-neutral pricing as a particular case and extends it to situations where risk-neutral pricing is impossible.
The second model is the Black-Scholes model for the equity index, where the pricing of contracts is performed within the risk-neutral framework. Crucially, we demonstrate that when the insurer prices and reserves using the Black-Scholes model, while the insured employs a dynamic withdrawal strategy based on the minimal market model, the insurer may be underestimating the value and associated reserves of the contract.



\end{abstract}

\end{frontmatter}


\section{Introduction}
Variable annuities (VA) with guarantees of living and death benefits are offered by wealth management and insurance companies worldwide to assist individuals in managing their pre-retirement and post-retirement financial plans. These products take advantage of market growth while providing a protection of the savings against market downturns. The VA contract cash flows received by the policyholder are linked to the choice of investment portfolio  (e.g. the choice of mutual fund and its strategy) and its performance while traditional annuities provide a pre-defined income stream in exchange for a lump sum payment. 

A variety of VA guarantees can be added by policyholders at the cost of additional fees. Common examples of VA guarantees include guaranteed minimum accumulation benefit (GMAB), guaranteed minimum withdrawal benefit (GMWB), guaranteed minimum income benefit (GMIB) and guaranteed minimum death benefit (GMDB), as well as combinations of these, e.g., guaranteed minimum withdrawal and death benefit (GMWDB), among others. These guarantees, generically denoted as GMxB, provide different
types of protection against market downturns, shortfall of savings due to longevity risk or assurance of stability of income streams. Precise specifications of these products can vary across categories and issuers; see \citet{bauerKR2008, Ledlie2008} and \citet{Kalberer2009} for an overview of these products.

Since the recent financial crisis, the need for accurate estimation of hedging costs of VA guarantees has become increasingly important. Such estimation includes the pricing of future cash flows that must be paid by the insurer to the policyholder in order to fulfill the liabilities of the VA guarantees, as well as the associated hedging strategy to deliver the liability payments. The standard pricing approach follows the risk-neutral pricing theory, where under the condition of no-arbitrage, the VA product is priced as the expectation of the totality of discounted future cash flows offered by the product under the so-called equivalent risk-neutral pricing measure. 
There have been a number of contributions in the academic literature considering numerical methods for the pricing of VA guarantees. These include standard and regression-based Monte Carlo, partial differential equation (PDE) and direct integration methods. A comprehensive overview of numerical methods for the pricing of VA guarantees is provided in \citet{ShevLuo2016}.

The benchmark approach (BA) offers an alternative pricing theory. 
Under very general conditions, there exists in a given investment universe a unique growth-optimal portfolio (GP). The BA takes the GP as the numeraire, or benchmark, such that any benchmarked nonnegative portfolio price process assumes zero expected instantaneous returns. The GP assumes the highest expected instantaneous growth rate among all nonnegative portfolios in the investment universe, and maximizes the expected log-utility of the terminal wealth. It is a well-diversified portfolio that draws on all tradable risk factors and the corresponding risk premia to achieve the growth optimality. Following the real-world pricing formula under the BA, real-world pricing of any given future cash flows identifies their minimal possible replication cost. The BA is considered as a generalization of the risk-neutral pricing theory, in that an equivalent risk-neutral pricing measure need not exist, yet it includes risk-neutral pricing as a special case involving additional conditions that ensure the existence of an equivalent risk-neutral probability measure. Another special case is the minimal market model (MMM), described in \ref{models}, which permits mean-reversion of the GP around an exponentially growing market trend and which captures both the leptokurtic behavior of logreturns, as shown in \citet{FergussonPl06dc}, and the leverage effect, manifest as a spike in volatility in a rapidly falling market; see for example \citet{campbell2005} and \citet{shiller2015}. In contrast, the Black-Scholes model (BSM), also described in \ref{models}, captures only mesokurtic behavior of logreturns, and is unable to model the leverage effect because of its constant volatility.

In this paper we consider a standard VA product with GMWB, which provides a guaranteed withdrawal amount per year until the maturity of the contract, regardless of the investment performance. The total withdrawal amount is such that the initial investment is guaranteed to be returned over the life of the contract. Additional features, such as a death benefit, can be added straightforwardly if so desired, see, e.g., \citet{luo2015}. 

Two classes of withdrawal strategies of the policyholder are often considered in the literature: a static withdrawal strategy under which the policyholder withdraws a predetermined amount on each withdrawal date; or a dynamic strategy where the policyholder ``optimally'' decides the amount of withdrawal at each withdrawal date depending on the information available at that date, where the optimality usually refers to the maximization of the value of the current and future cash flows. By assuming an optimal policyholder's withdrawal behavior, the pricing of the VA product corresponds to the hedging cost of the worst case scenario faced by the VA provider. In other words, the price of the VA product under the respective dynamic strategy provides an upper bound of hedging cost from the VA provider's perspective; see \citet{Sun2018}. It should be noted that the actual policyholders' withdrawal strategies could be far from optimal; see, e.g., \citet{MoenigB2015}.

Assuming that the policyholder takes the dynamic withdrawal strategy, that is, the optimal strategy that maximizes the present value of current and future cash flows of the VA product, the actual withdrawals still depend on the pricing method adopted by the policyholder. On the other hand, the VA provider, who maintains a hedging portfolio to deliver the liability cash flows of the VA product also faces the same choices for pricing and hedging for the portfolio. In this paper, we consider two pricing methods, the risk-neutral pricing approach and the BA. We investigate the outcomes and implications of different choices of withdrawal and hedging strategies by the policyholder and the VA provider. In particular, we study empirically the cases when the two parties take different pricing approaches. 
In the VA pricing literature it is most often the case that the same pricing model is adopted by both the policyholder and the VA provider. The important situation where the policyholder and the VA provider hold fundamentally different valuation perspectives, as is described in this paper, has not been investigated. This paper attempts to fill this vacancy and hopefully initiate more interest in this direction.

The paper is organized as follows. In Section \ref{va} we present the contract details of the GMWB guarantee together with its pricing formulation under a stochastic optimal control framework. 
Then, in Section \ref{num} we empirically test these pricing and corresponding withdrawal and hedging strategies, respectively, for the policyholder and the VA provider, when the two parties use the same or different modeling approaches. Section \ref{conc} concludes with remarks and discussion.
The appendices contain background technical aspects of our VA modeling and pricing approaches.
\ref{ba} provides an overview of the benchmark approach.
In \ref{models} we describe the pricing models under the risk-neutral approach and the BA. 
Finally, in \ref{pricing} the VA pricing problem is formulated for optimal policyholder's withdrawals under both pricing frameworks and models, and the numerical algorithm to solve the problem is described. 

\section{Description of the VA Guarantee Product}
\label{va}
We consider the VA product where a policyholder invests at time $t=0$ a lump-sum of $W(0)$ into a wealth account $W(t),t\in[0,T]$ that tracks an equity index $S(t), t\in[0,T]$, where $t=T$ corresponds to the expiry date of the VA contract. We assume both $W(t)$ and $S(t)$ are discounted by the locally risk-free savings account, as are all other values of wealth encountered hereafter. The (discounted) equity index evolves under the real-world probability measure $\Pmeas$ according to the SDE
\beql{index}
\frac{dS(t)}{S(t)}=\mu(t)dt+\sigma(t)dB(t),
\qquad t\in[0,T],
\eeq
where $\mu(t)$ and $\sigma(t)$ are the instantaneous market risk premium and volatility of the index, respectively. Here $B(t)$, $t\in[0,T]$, is a standard $\Pmeas$-Brownian motion driving traded market uncertainties. 

As mentioned in \ref{models}, the diversified equity index approximates well the GP, and we have, as a particular case of \eqref{index}, 
the Black-Scholes model (BSM) of the GP, specified by the SDE \eqref{Eqn:BS}.
Also, as another particular case of \eqref{index}, we have the minimal market model (MMM) of the GP, specified by the SDE \eqref{smm0}.
Further details of both of these models are supplied in \ref{models}.

The policyholder selects a GMWB rider in order to protect his wealth account $W(t)$, $t\in[0,T]$, over the lifetime of the VA contract. 
The GMWB contract allows the policyholder to withdraw from a guarantee account $A(t)$, $t\in[0,T]$, on a sequence of pre-determined contract event dates, $0=t_0<t_1<\dots<t_N=T$. The initial guarantee $A(0)$ matches the initial wealth $W(0)$. We assume here that the guarantee account stays constant over time, unless a withdrawal is made on one of the event dates, which reduces the guarantee account balance. Other forms of guaranteed returns can be modeled similarly.
For simplicity, we do not include in our discussion features such as death or early surrender benefits. Under a more realistic setting, these additional features can be included straightforwardly within the framework described in this paper.

To simplify notations, we denote by $\bX(t)$ the vector of state variables at time $t$, given by
\beql{statev}
\bX(t)=(\mu(t),\sigma(t),S(t),W(t),A(t)),\quad t\in[0,T].
\eeq
Here, we assume that the variable $S(t)$ follows a Markov process, so that $\bX(t)$ contains all the market and account balance information available at $t$. For simplicity, we assume that the state variable $S(t)$ is continuous, and $W(t)$ and $A(t)$ are left-continuous with right-hand limit (LCRL). 

On event dates $\tn,n=1,\dots,N$, a nominal withdrawal $\gamma_n$ from the guarantee account is made. The policyholder may choose $\gamma_n\le A(\tn)$ on $\tn<T$. Otherwise, a liquidation withdrawal of $\max(W(\tn),A(\tn))$ is made. That is, 
\beql{gamman}
\gamma_n=
\begin{cases}
\Gamma(\tn,\bX(\tn)), & {n<N},\\
\max\Prt{W(\tn),A(\tn)}, & {n=N},
\end{cases}
\eeq
where $\Gamma(\cdot,\cdot)$ is referred to as the \emph{withdrawal strategy} of the policyholder.
The net cash flow received by the policyholder, which may differ from the gross amount, is denoted by $C_n(\gamma_n,\bX(\tn))$. In our case this cash flow is set to
\beql{cfd}
C_n(\gamma_n,\bX(\tn))=
\begin{cases}
\gamma_n-\beta\max(\gamma_n-G_n,0), & n<N,\\
\max\Prt{W(T),A(T)}-\beta\max(A(T)-G_N,0), & n=N,
\end{cases}
\eeq
where $G_n$ is a pre-determined withdrawal amount specified in the GMWB contract, and $\beta$ is the penalty rate applied to the part of the withdrawal from the guarantee account exceeding the contractual withdrawal $G_n$. Here we assume the penalty also applies to the last withdrawal of the guarantee account $A(T)$. The part of the wealth account balance in excess of the guarantee account balance is not subject to the penalty. 
Upon withdrawal by the policyholder, the guarantee account is reduced by the nominal withdrawal $\gamma_n$, that is, 
\beql{ajump}
A(\tn^+)=A(\tn)-\gamma_n,
\eeq 
where $A(\tn^+)$ denotes the guarantee account balance ``immediately after" the withdrawal. Note that $A(\tp)\ge0$. The wealth account is reduced by the amount $\min(\gamma_n,W(\tn))$ and remains nonnegative. That is,
\beql{wjump}
W(\tn^+)=\max(W(\tn)-\gamma_n,0),
\eeq
where $W(\tn^+)$ is the wealth account balance immediately after the withdrawal. It is assumed that $\gamma_0=0$, i.e., there are no withdrawals at the start of the contract. Both the wealth and the guarantee account balance are 0 after contract expiration. That is, we have
\beql{WAterm}
W(T^+)=A(T^+)=0.
\eeq

Throughout the VA contract, the wealth account is charged an insurance fee continuously at rate $\alpha_{\rm tot}$ for the GMWB rider by the insurer to pay for the hedging cost of the guarantee. Discrete fees may be modeled similarly without any difficulty. The wealth account in turn evolves as
\beql{wsde}
\frac{dW(t)}{W(t)}=(\mu(t)-\alpha_{\rm tot}(t))dt+\sigma(t)dB(t),
\eeq
for any $t\in[0,T]$ at which no withdrawal of wealth is made. Here, $\alpha_{\rm tot}(t)=\alpha_{\rm ins}(t)+\alpha_{\rm m}(t)$ is the total fee rate, where $\alpha_{\rm ins}$ denotes the insurance fee and $\alpha_{\rm m}$ denotes the management fee. 

We denote the VA \emph{plus} guarantee value function at time $t$ by $V(t,\bX(t)),t\in[0,T]$, which corresponds to the present value under the respective pricing model of all future cash flows entitled to the policyholder on or after the current time $t$. The remaining value after the final cash flow is, obviously, 0, i.e.,
\beql{Vterm}
V(T^+,\bX(T^+))=0.
\eeq

\section{Backtesting the Reserving and Policyholder Strategies}
\label{num}
In this section, we conduct backtests of the two proposed pricing models for VA products. We consider both from the policyholder's perspective, where he or she decides the optimal withdrawal amount based on the pricing model chosen by the policyholder, and from the VA provider's perspective, where the strategy of the hedging portfolio for the VA product is based on the pricing model chosen by the provider. 

We take the S\&P500 index as the equity index underlying the VA product, and run simulated withdrawals and hedging strategies on the historical data of the underlying well-diversified index. In particular, we take the historically observed monthly prices of the S\&P500 index from 1871 to 2018, with all dividends reinvested, and discounted by the locally risk-free savings account. The S\&P500 data after 1963 are obtained from Datastream, and the earlier part from 1871 until 1963 is reconstructed in \citet{shiller2015}. The S\&P500 total return index provides a good approximation of the dynamics of the market portfolio (MP) of the US domestic stock market. We  consider  the pricing of a GMWB contract written on a VA account tracking the index over the 30-year period from Feb. 1988 to Feb. 2018, based on the underlying models estimated from the historical index prices prior to this period.  We consider both the MMM and the BSM as the underlying dynamics, as specified by SDEs \eqref{smm0} and \eqref{Eqn:BS} respectively, and compare the evolution of the guarantee values under both models and respective pricing rules.

\begin{figure}[htp!]
\centering
\begin{tabular}{cc}
\includegraphics[width=7cm]{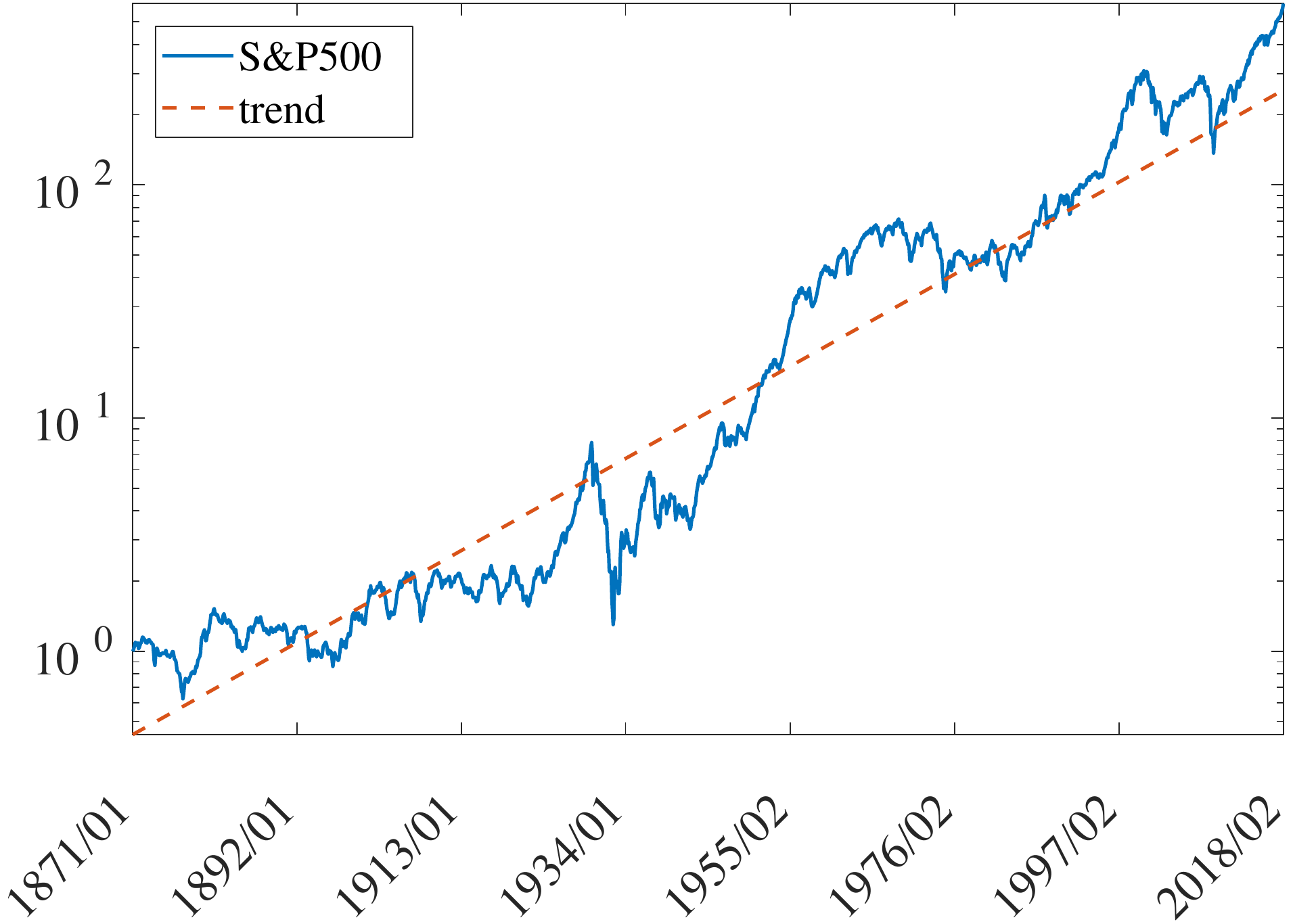}&
\includegraphics[width=7cm]{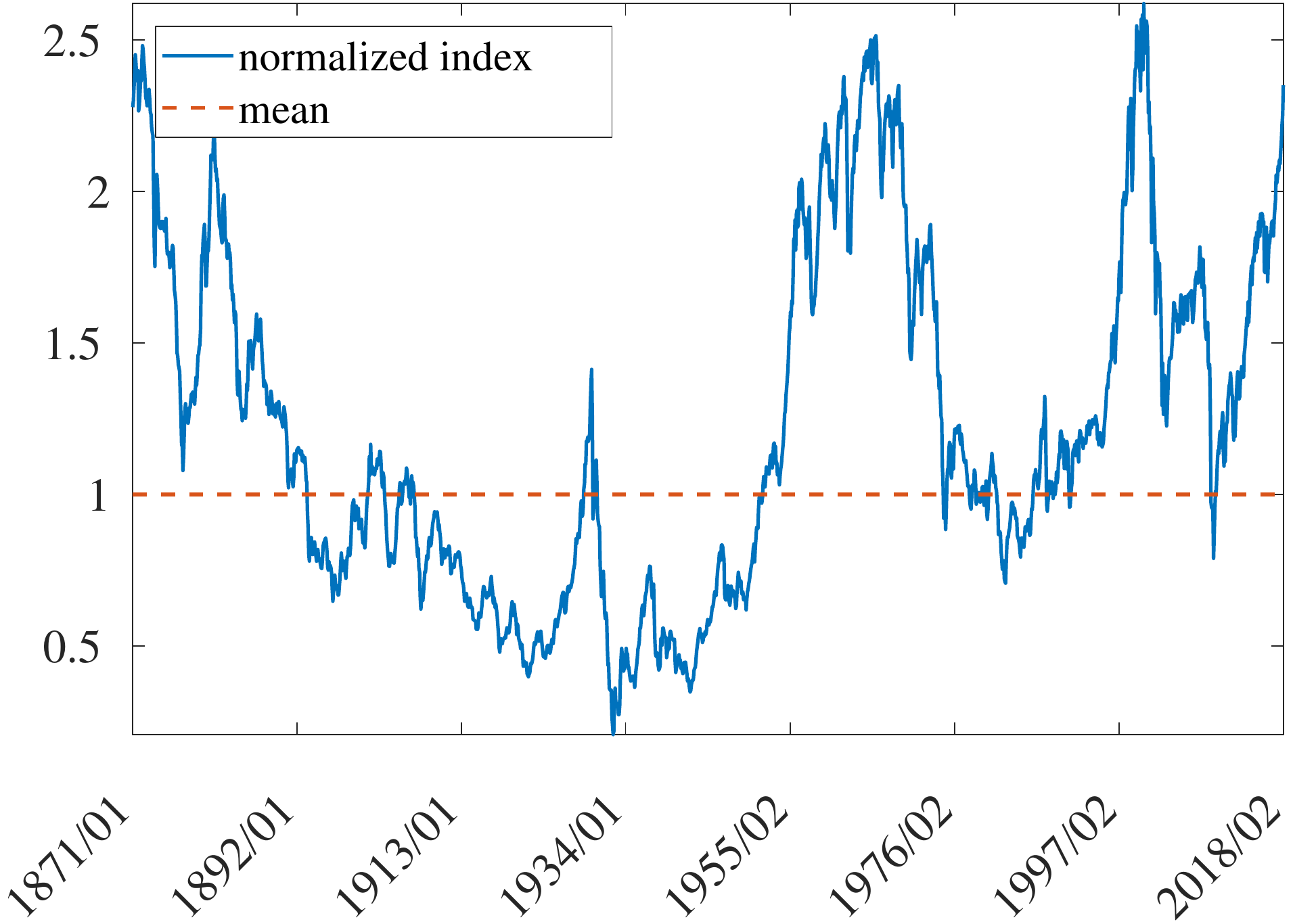}\\
(a) S\&P500 index with the estimated trend & (b) normalized index with the mean\\
\end{tabular}
\caption{Estimated trend of growth and normalized index for the MMM.}
\label{sp500log}
\end{figure}

The log-prices of the S\&P500 total return index are shown in Figure \ref{sp500log}~(a). We take the first 97 years of the available data, from Jan. 1871 to Feb 1988 for estimations.  The MMM parameters were estimated as follows:
To estimate the overall trend of growth $\frac{S(t)}{Y(t)}=\frac{\alpha_0}{\eta}e^{\eta t}$ for the MMM, a straight line is fitted to the log-index prices and the slope is taken as the estimated total growth rate $\eta$, and the scaling factor $\alpha_0$ is determined from the intercept. The normalized index $Y(t)$ thus obtained is shown in Figure \ref{sp500log}~(b). The BSM volatility in \eqref{Eqn:BS} was estimated following the standard MLE estimator. The estimated parameters are shown in Table~\ref{estimates}. 

\begin{table}
\caption{Estimated model parameters from the S\&P500 historical prices.}
\centering
\begin{tabular}{r|ccccc}
	\toprule
	& $\alpha_0$ & $\eta$  & $\sigma$ \\ 
	\midrule
	MMM & 2.857 & 0.0435 & -\\ 
	BSM & - & -  & 0.1441 \\ 
	\toprule
\end{tabular}
\label{estimates}
\end{table}

We consider a stylized VA contract with GMWB where the policyholder invests 1 Million units of the US dollar savings account in a mutual fund that tracks the S\&P500 total return index. For illustrative purposes, we assume there are no mutual fund management fees, so that $\alpha_{\rm m}(t)=0$ in \eqref{wsde}, and the insurance fee rate $\alpha_{ins}$ is set at $0$. (The case with nonzero total fees can be considered strictly analogously without affecting the main results from the current discussion. )
The policyholder purchases a GMWB rider that guarantees equal annual payments of the initial investment of 1 Million savings account units over a period of 30 years. The contracted withdrawals are, therefore, rated at 33,333 units of the savings account per annum. If the policyholder decides to withdraw more than the contracted amount, a penalty charge of 10\% should apply to the excess part of the withdrawal. As mentioned in Section \ref{va}, we assume that the penalty charge also applies to thew last withdrawal. That is, if the balance of the guarantee account exceeds the contracted withdrawal amount at maturity, withdrawal of this balance is mandatory and the same penalty rate applies to the excess part. 

We first consider the situation where the VA provider prices the product under the BSM with the risk-neutral pricing approach (see \ref{models}) assuming that the policyholder makes optimal withdrawals under the same pricing rule. The VA provider maintains a nominal wealth account $W(t)$ of the policyholder's wealth, and a nominal guarantee account $A(t)$ to keep track of the remaining guaranteed withdrawal allowance. The VA provider maintains an actual hedging portfolio $V(t)$ consisting of shares of the index-tracking mutual fund, or the index for short, and the locally risk-free savings account. We refer to the hedging portfolio as the reserve account of the VA product, which is the only real investment account involved. The reserve account starts at value $V(0)$, the initial price of the VA product, and maintains a self-financing hedging strategy until a withdrawal is made on one of the withdrawal dates $\tn$, when the net cash flow $C_n$ is paid out of this account to the policyholder. The strategy maintained by the reserve account between withdrawal dates is the respective delta-hedging strategy. 

\begin{figure}[htp!]
	\centering
	\begin{tabular}{cc}
		\includegraphics[width=6cm]{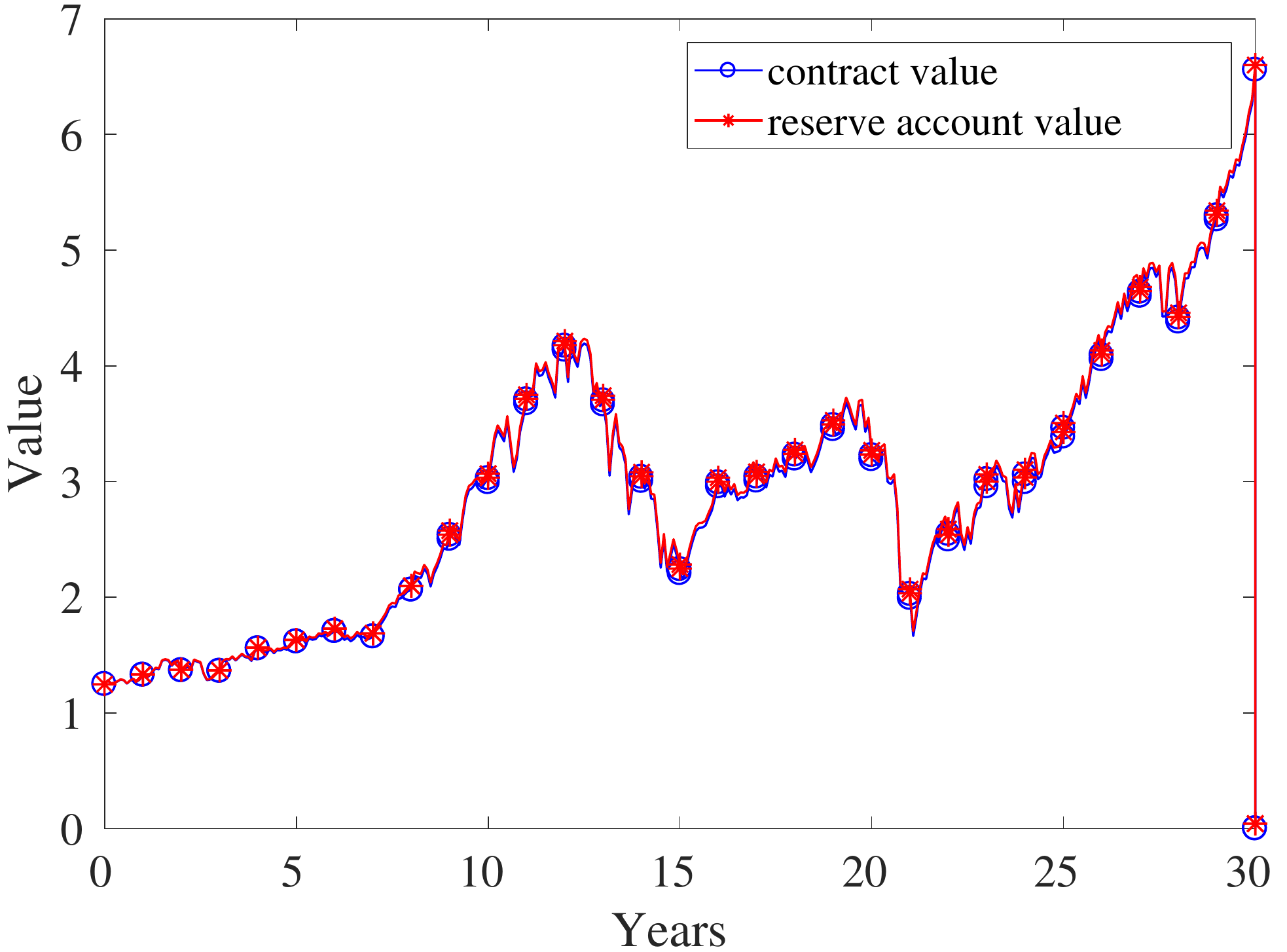}&
		\includegraphics[width=6cm]{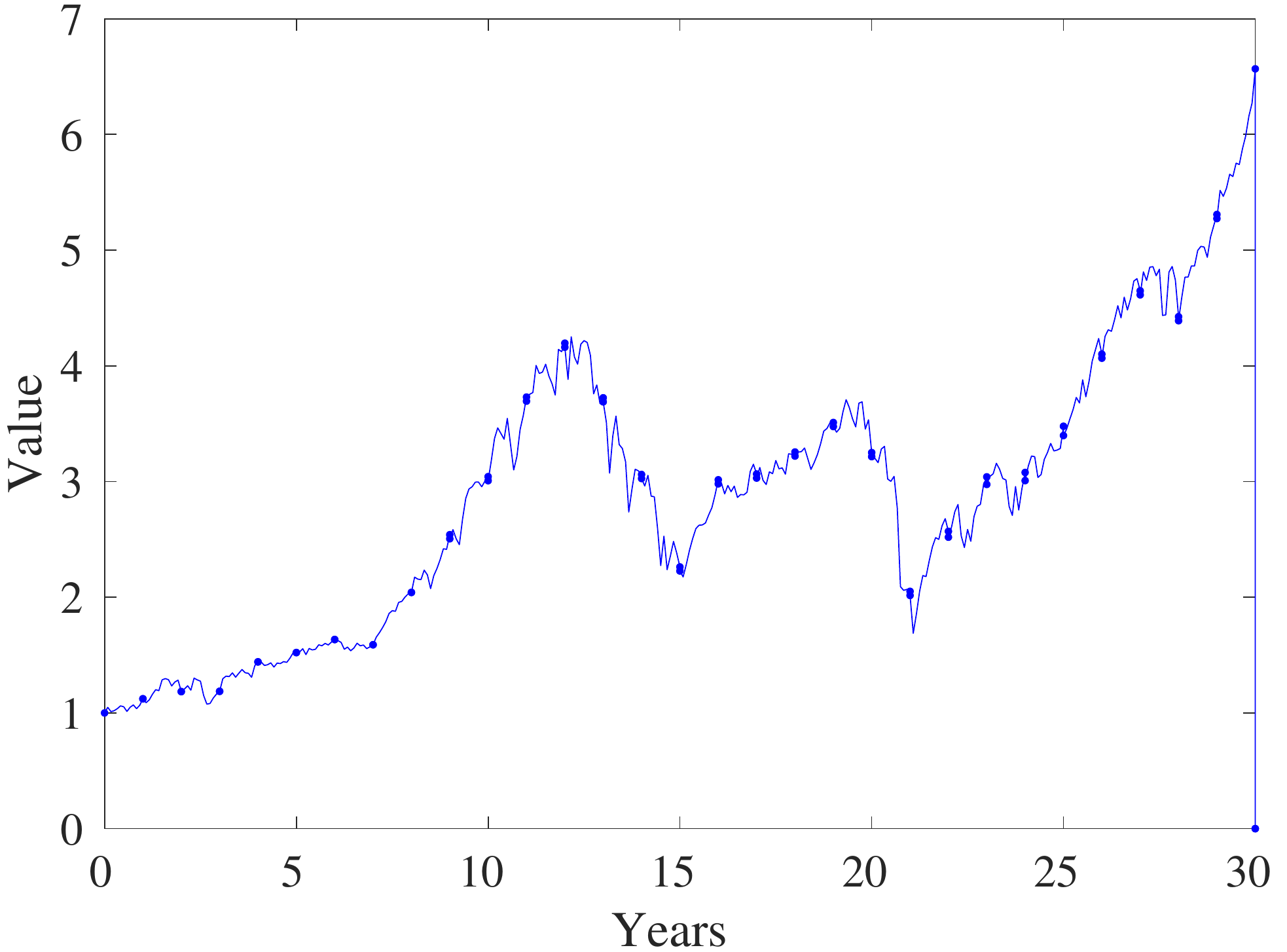}\\
		(a) contract value and reserve account  & 
		(b) nominal wealth account \\
		\includegraphics[width=6cm]{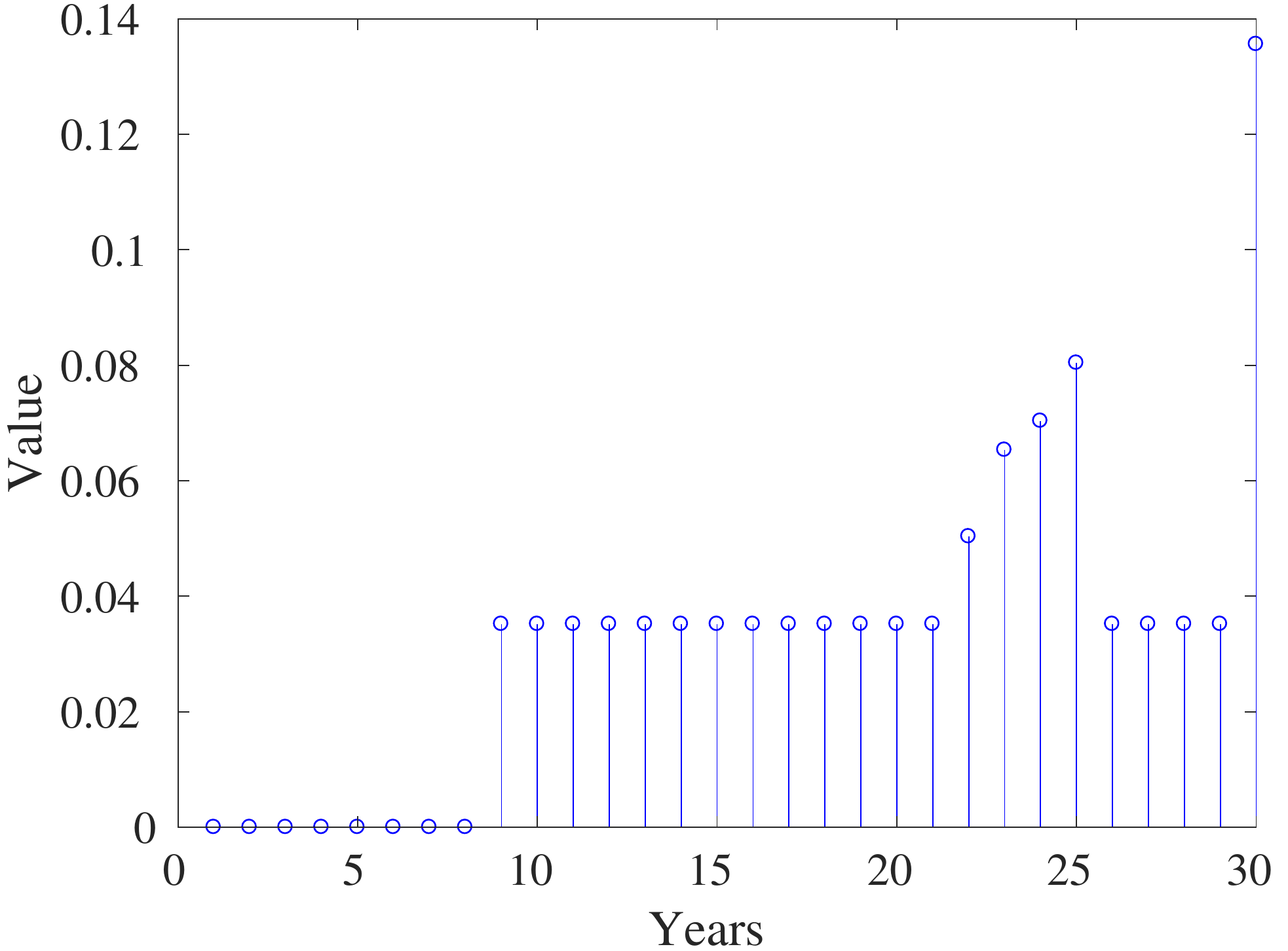}&
		\includegraphics[width=6cm]{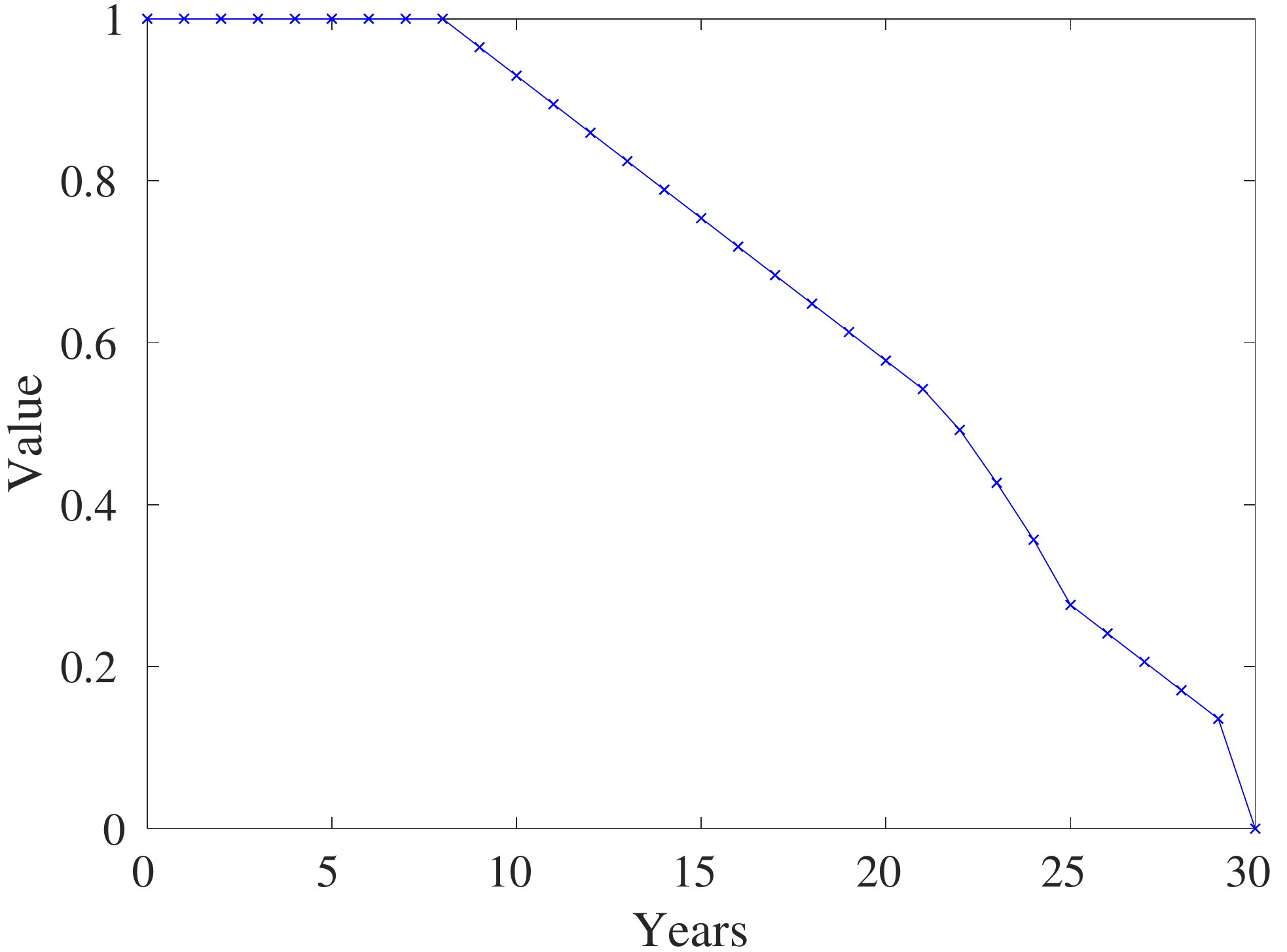}\\
		(c) withdrawals & 
		(d) nominal guarantee account \\
	\end{tabular}
	\caption{Value processes (in millions of units of the savings account) associated with the VA product when the pricing and hedging as well as optimal withdrawals are performed based on the BSM and risk-neutral pricing.}
	\label{vbsfig}
\end{figure}

Following Algorithm \ref{algo1} described in \ref{pricing}, we compute recursively the price process of the VA product, based on the historical index prices. The price process is shown in Figure~\ref{vbsfig}~(a), with an initial value of 1.22 Million, and a terminal value of 7.11 Million before the final liquidation. The reserve account process, realized through delta-hedging, is shown in the same plot for comparison. The intial value of the reserve account is the same as the price process, and the terminal value is 7.07 Million. After liquidation, the reserve account ended up with a small deficit of -0.0344 Million, possibly due to hedging errors from discrete hedging. 

The nominal wealth account $W(t)$ is shown in Figure~\ref{vbsfig}~(b),   
the optimal withdrawals made by the policyholder are shown in  Figure~\ref{vbsfig}~(c), and the guarantee account balance is shown in Figure~\ref{vbsfig}~(d). Note that both the wealth account and the guarantee account are nominal only, used for keeping track of the status of the policyholder's VA contract. No actual trading happens to these accounts. The optimal withdrawals are relatively uniform, except for no withdrawals in the beginning periods, and a large withdrawal in the last period. The relatively uniform withdrawal behavior is typical for the BSM and risk-neutral pricing, where the equity index dynamics is time-homogeneous. The only motivations to change the withdrawal pattern are changing the maturity date and wealth / guarantee account ratio. 

\begin{figure}[htp!]
	\centering
	\begin{tabular}{cc}
		\includegraphics[width=6cm]{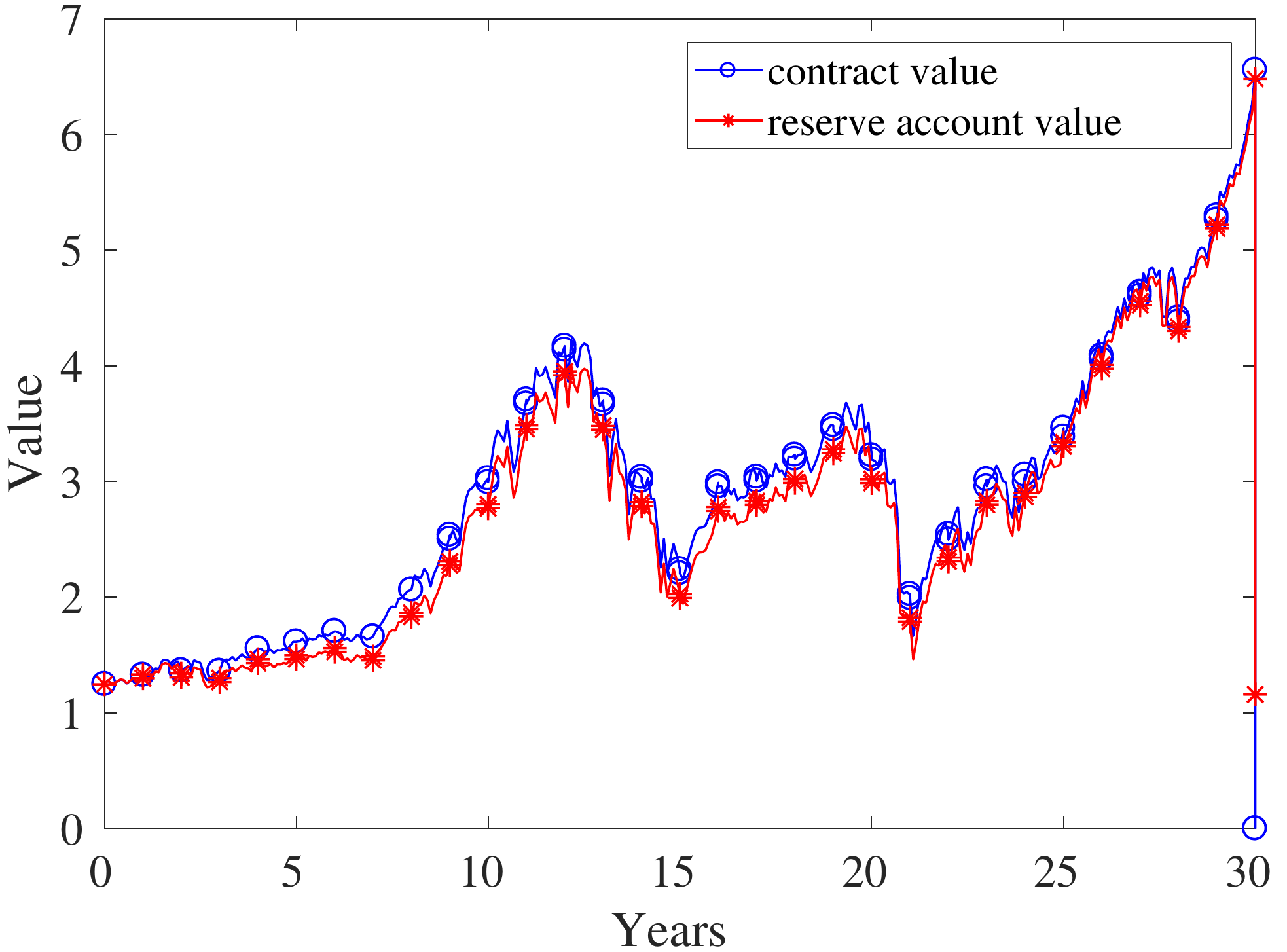}&
		\includegraphics[width=6cm]{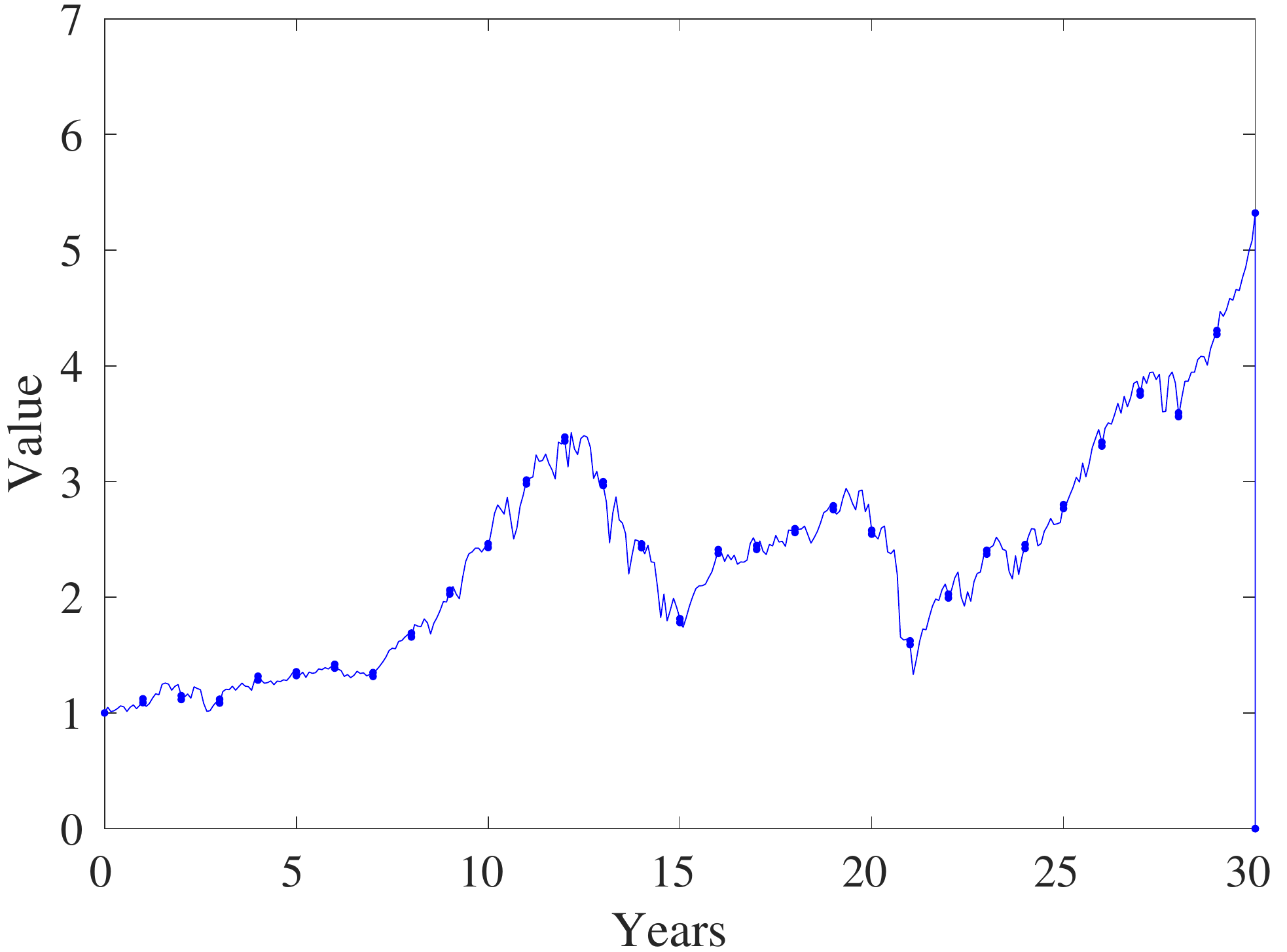}\\
		(a) contract value and reserve account  & 
		(b) nominal wealth account \\
		\includegraphics[width=6cm]{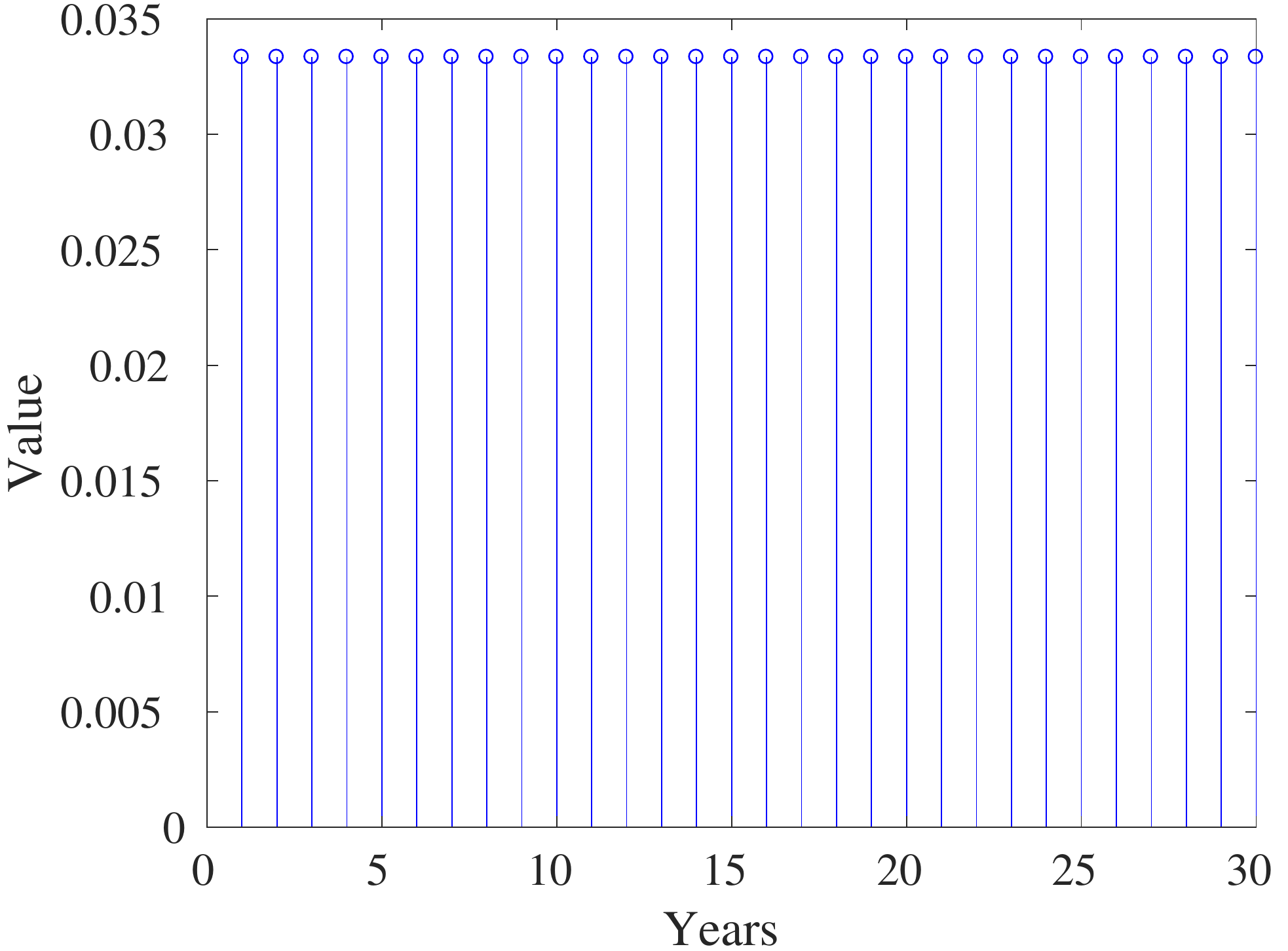}&
		\includegraphics[width=6cm]{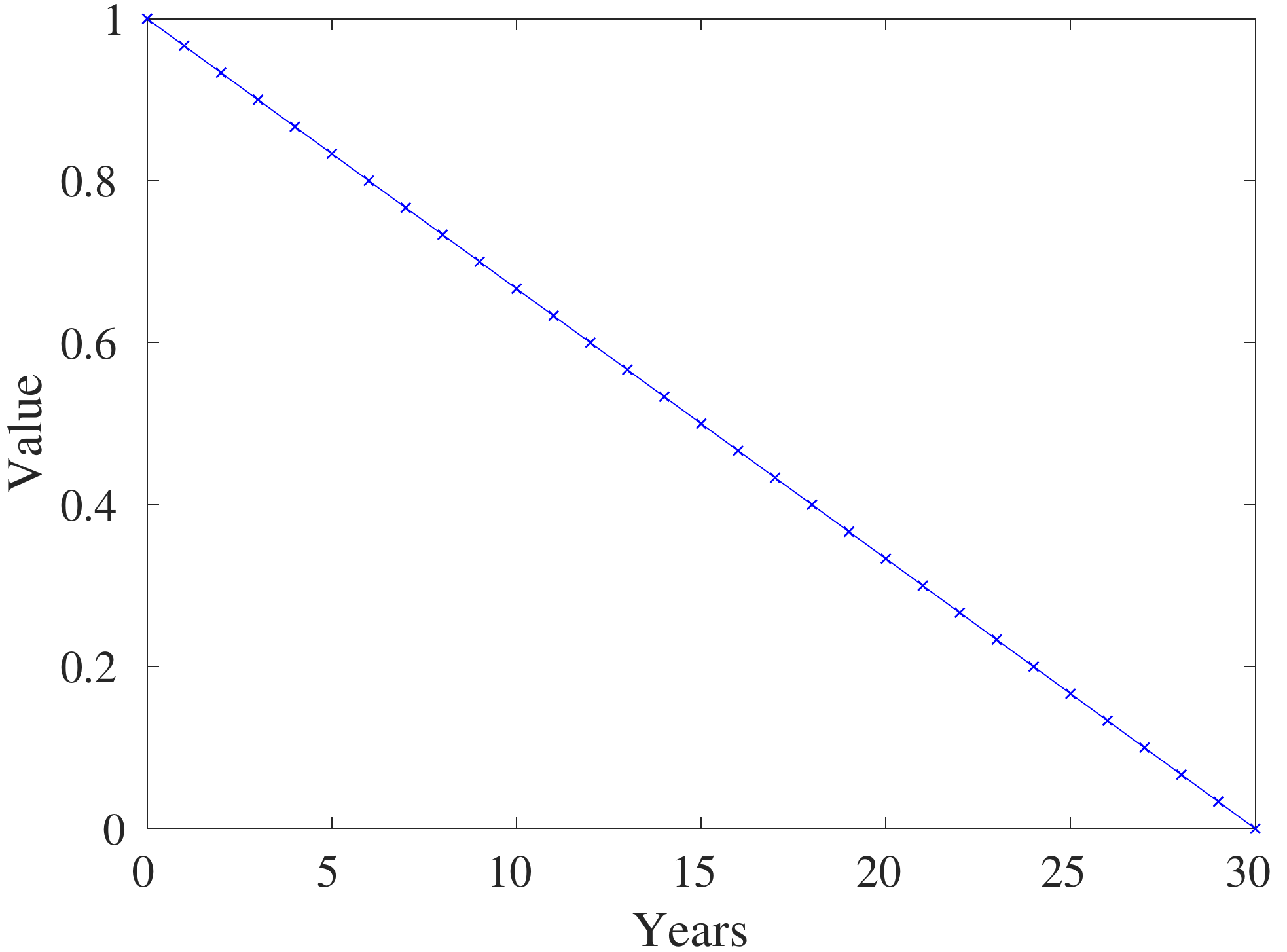}\\
		(c) withdrawals & 
		(d) nominal guarantee account \\
	\end{tabular}
	\caption{Value processes associated with the VA product when the pricing and hedging are performed based on the BSM and risk-neutral pricing  assuming an optimal withdrawal behavior, while the actual withdrawal behavior follows the static strategy.}
	\label{sbsfig}
\end{figure}

For verification purposes we consider the alternative static withdrawal behavior from the policyholder. That is, we assume the policyholder makes a uniform withdrawal equal to $1/N$ Million units of the savings account on any one of the $N$ withdrawal dates. On the other hand, the VA provider manages the reserve account in the same way as in the optimal case irrespective of the policyholder's withdrawal behavior, which is considered suboptimal in this case. The results are shown in Figure~\ref{sbsfig}. The contract value process in this case is the same as for the case with optimal withdrawals. The reserve account process as well as the nominal wealth account process differ from the previous case due to different (suboptimal) withdrawals. The reserve account ended up with a rather significant surplus of 1.39 Million. This is due to the loss made by the policyholder for withdrawing suboptimally. In particular, premature withdrawals led to less wealth accumulations in the nominal wealth account, leading to significantly less liquidation cash flow entitled to the policyholder. Since the reserve account maintained the same hedging strategy as in the previous case, it ended up having a surplus after paying the reduced liabilities. 

\begin{figure}[htp!]
	\centering
	\begin{tabular}{cc}
		\includegraphics[width=6cm]{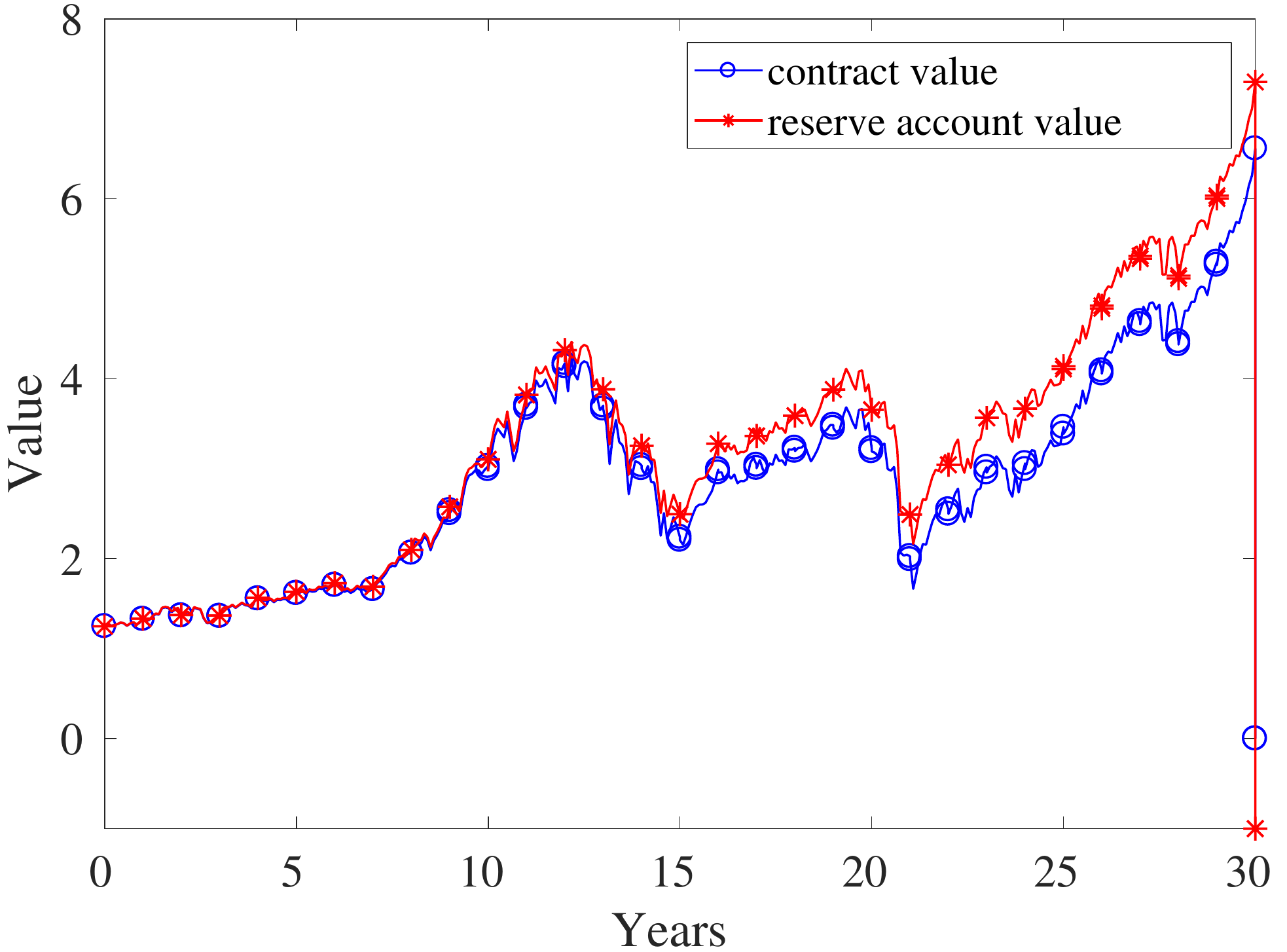}&
		\includegraphics[width=6cm]{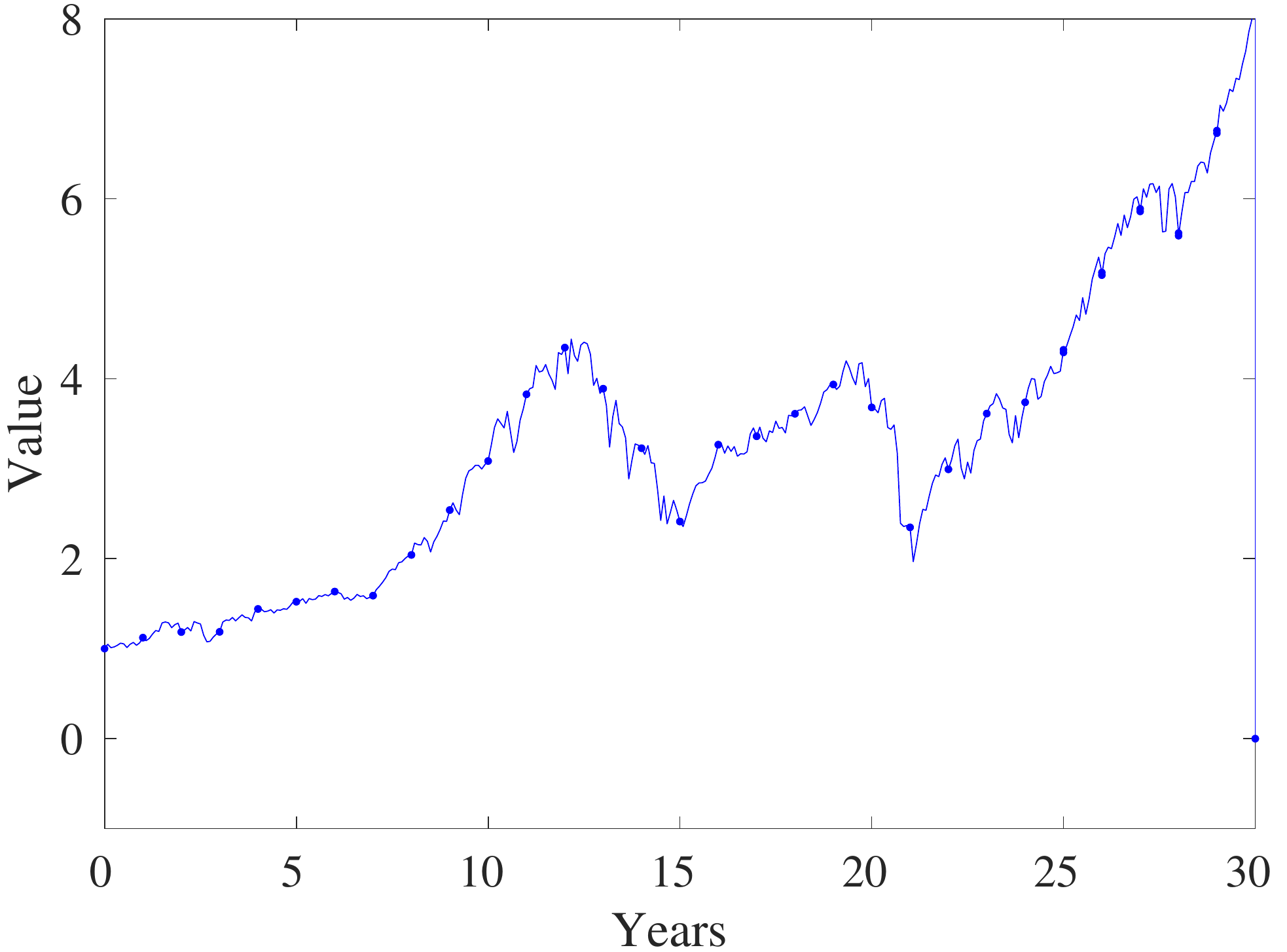}\\
		(a) contract value and reserve account  & 
		(b) nominal wealth account \\
		\includegraphics[width=6cm]{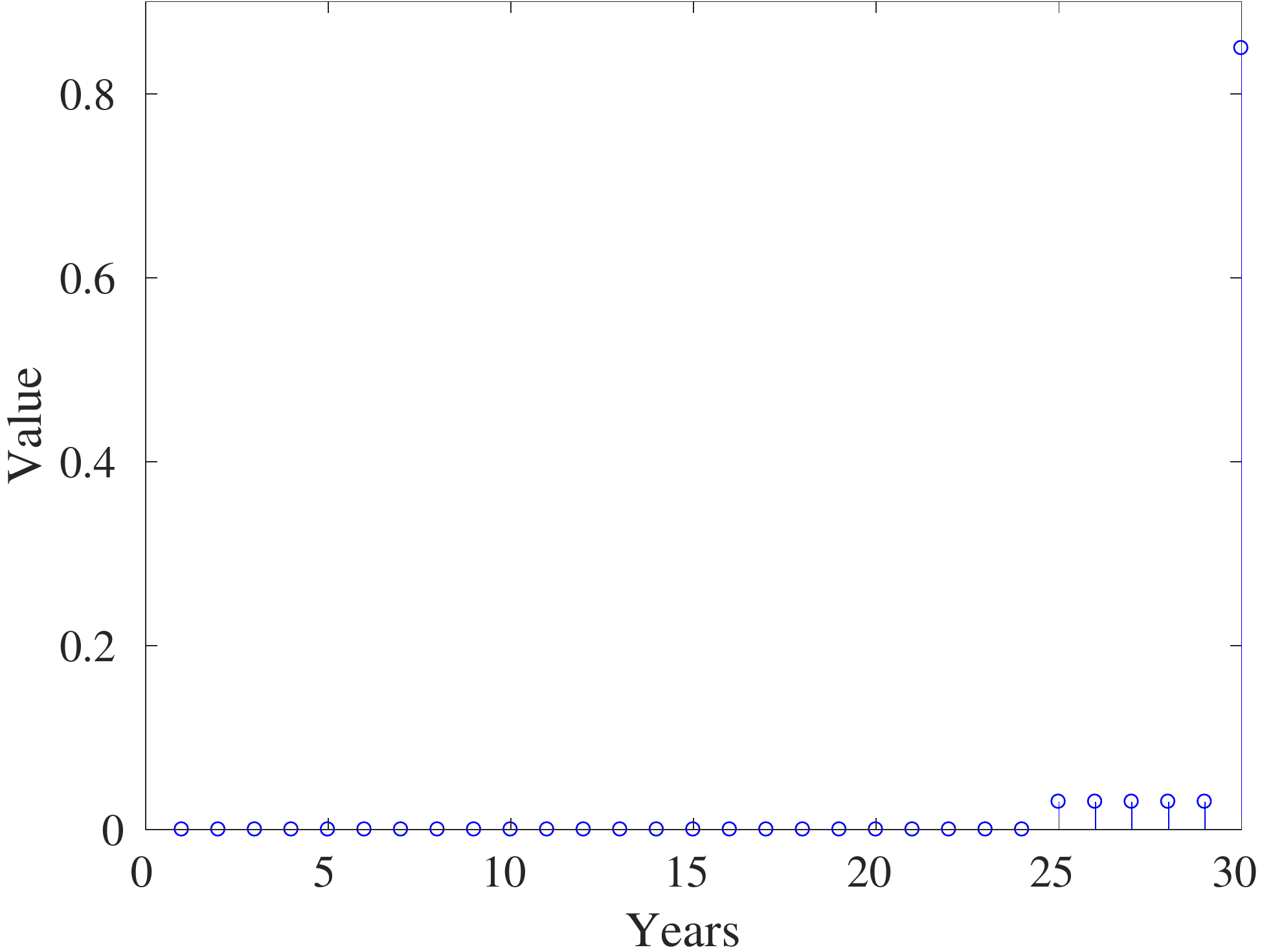}&
		\includegraphics[width=6cm]{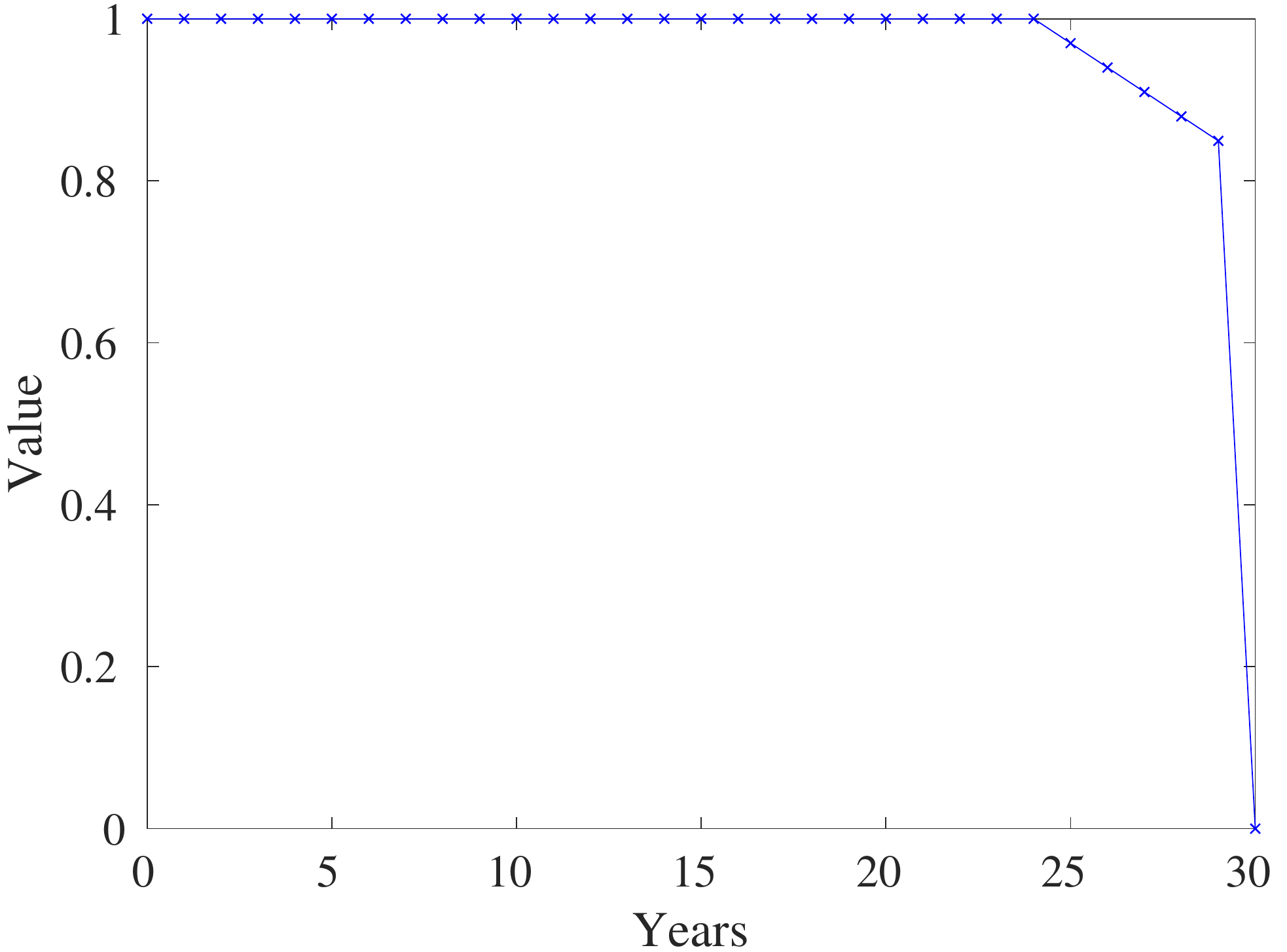}\\
		(c) withdrawals & 
		(d) nominal guarantee account \\
	\end{tabular}
	\caption{Value processes associated with the VA product when the pricing and hedging are performed based on the BSM and risk-neutral pricing  assuming an optimal withdrawal behavior, while the actual withdrawal behavior follows the MMM under the BA.}
	\label{mmbsfig}
\end{figure}

We next consider the situation where the policyholder makes withdrawals based on the MMM under the BA. That is, the policyholder's withdrawals maximize the value of the VA contract as priced by the MMM under the BA. The VA provider, believing in the BSM under the risk-neutral pricing framework, manages the reserve account in the same way as in the previous cases, and views the policyholder's withdrawals as being suboptimal. The VA provider thus expects to receive a surplus in the reserve account after maturity of the VA contract. The outcomes of this scenario are shown in Figure~\ref{mmbsfig}. 

To the surprise of the VA provider, instead of having a surplus, the reserve account in this case ended up with a deficit of 1 Million, as indicated in Figure~\ref{mmbsfig}~(a). The withdrawal behavior of the policyholder is such that there are no withdrawals until the very end of the contract term, where a number of small withdrawals were followed by a large withdrawal on the maturity date. The nominal wealth account accumulated a high level of wealth due to no withdrawals in the early stages. The liquidation of this large wealth led to the deficit of the reserve account, which followed a hedging strategy assuming more early withdrawals. 

The failure of the VA provider in hedging the VA product, when the policyholder behaved optimally under a different pricing framework, indicates the potential inappropriateness of the BSM and risk-neutral pricing adopted by the VA provider. In particular, the policyholder believed in the long-term growth of the market and invested for this growth according to the MMM. The VA provider, from a risk-neutral perspective, did not recognize the long-term growth, and managed the reserve account with a short-term vision, leading to the failure of matching the performance of the policyholder's wealth account. Note that the MMM and the associated long-term growth rate were estimated from prior returns of the index. Thus, no ``looking into the future" is associated with the policyholder's withdrawal behavior. 

\begin{figure}[htp!]
	\centering
	\begin{tabular}{cc}
		\includegraphics[width=6cm]{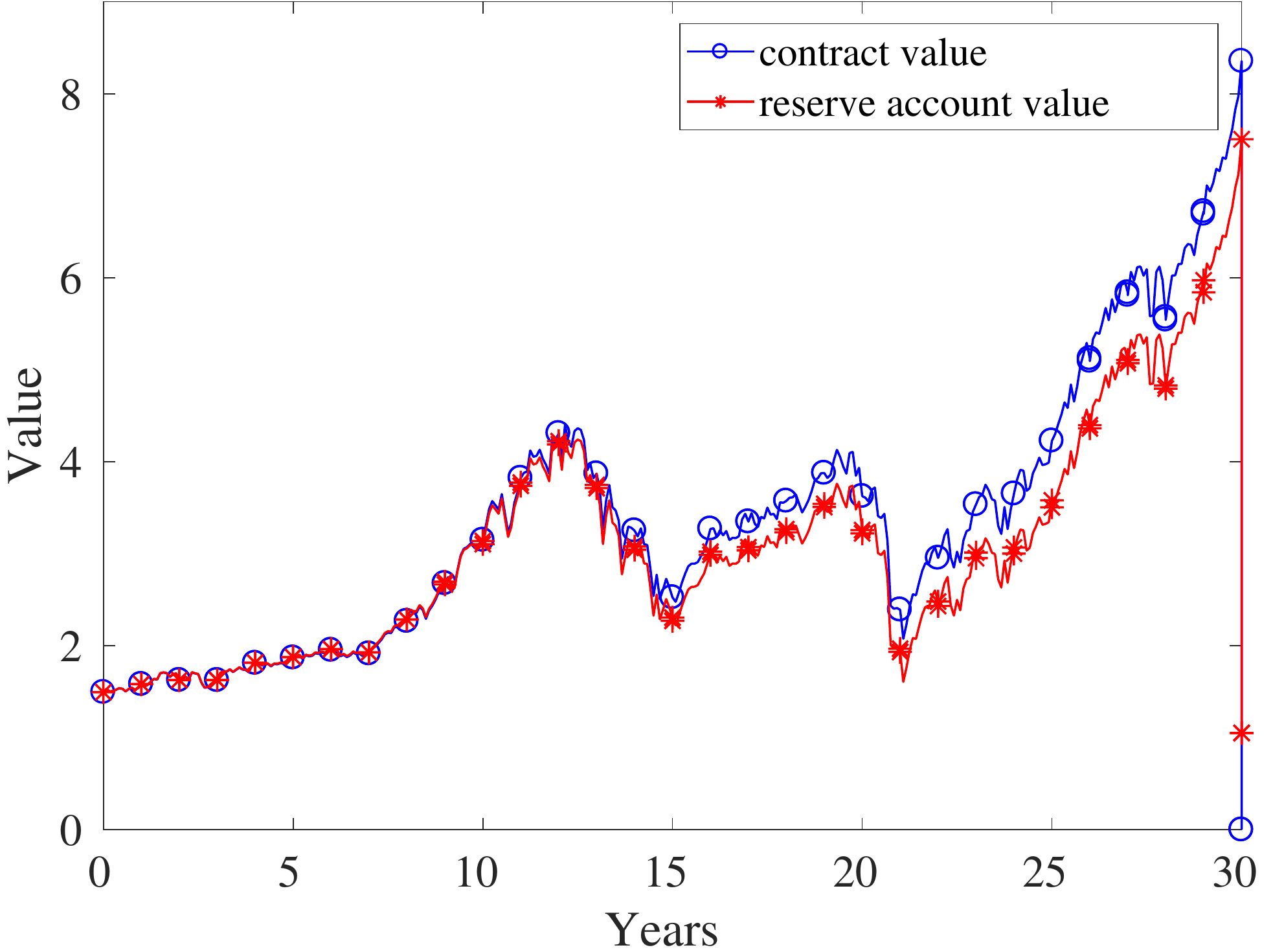}&
		\includegraphics[width=6cm]{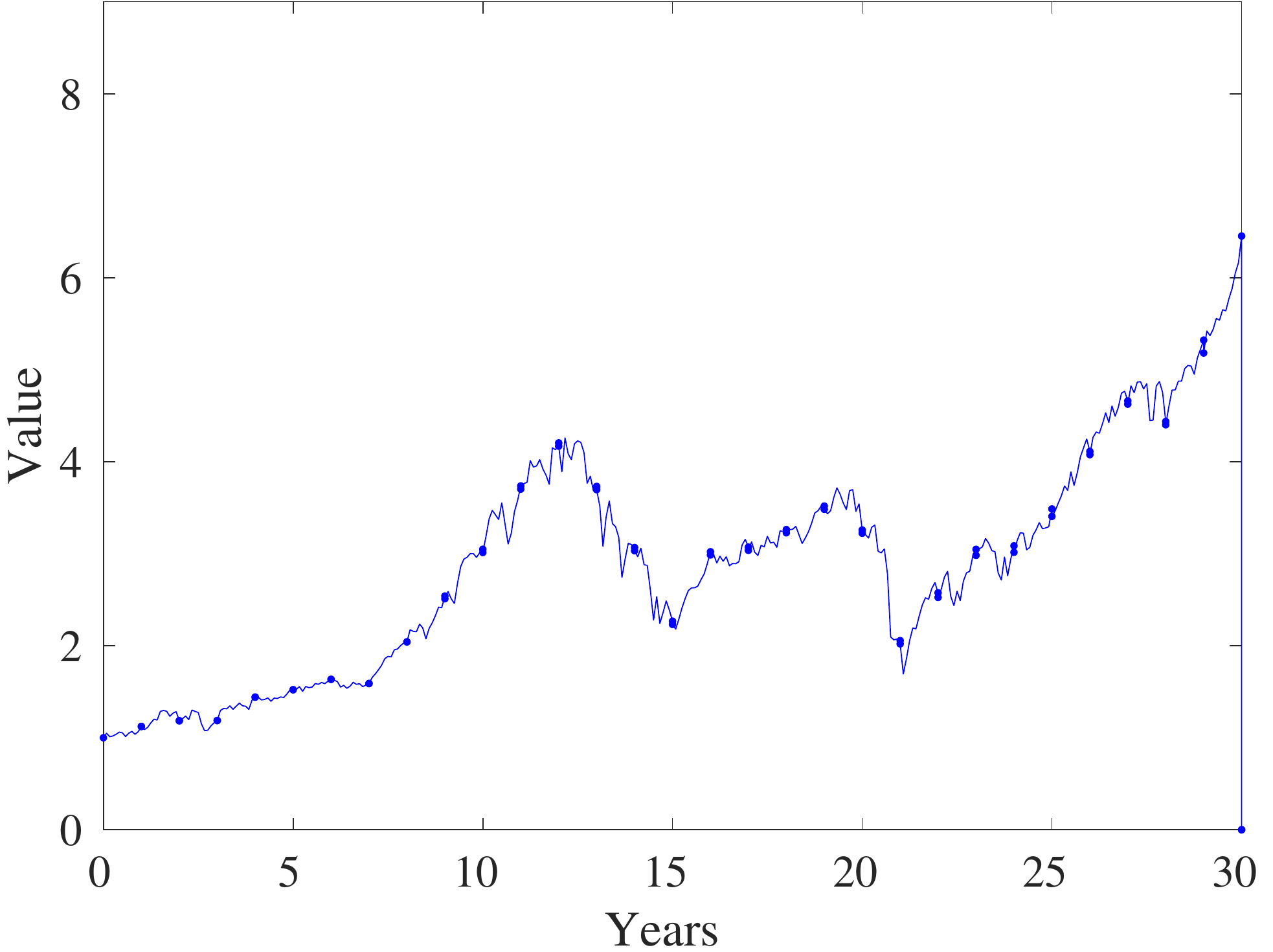}\\
		(a) contract value and reserve account  & 
		(b) nominal wealth account \\
		\includegraphics[width=6cm]{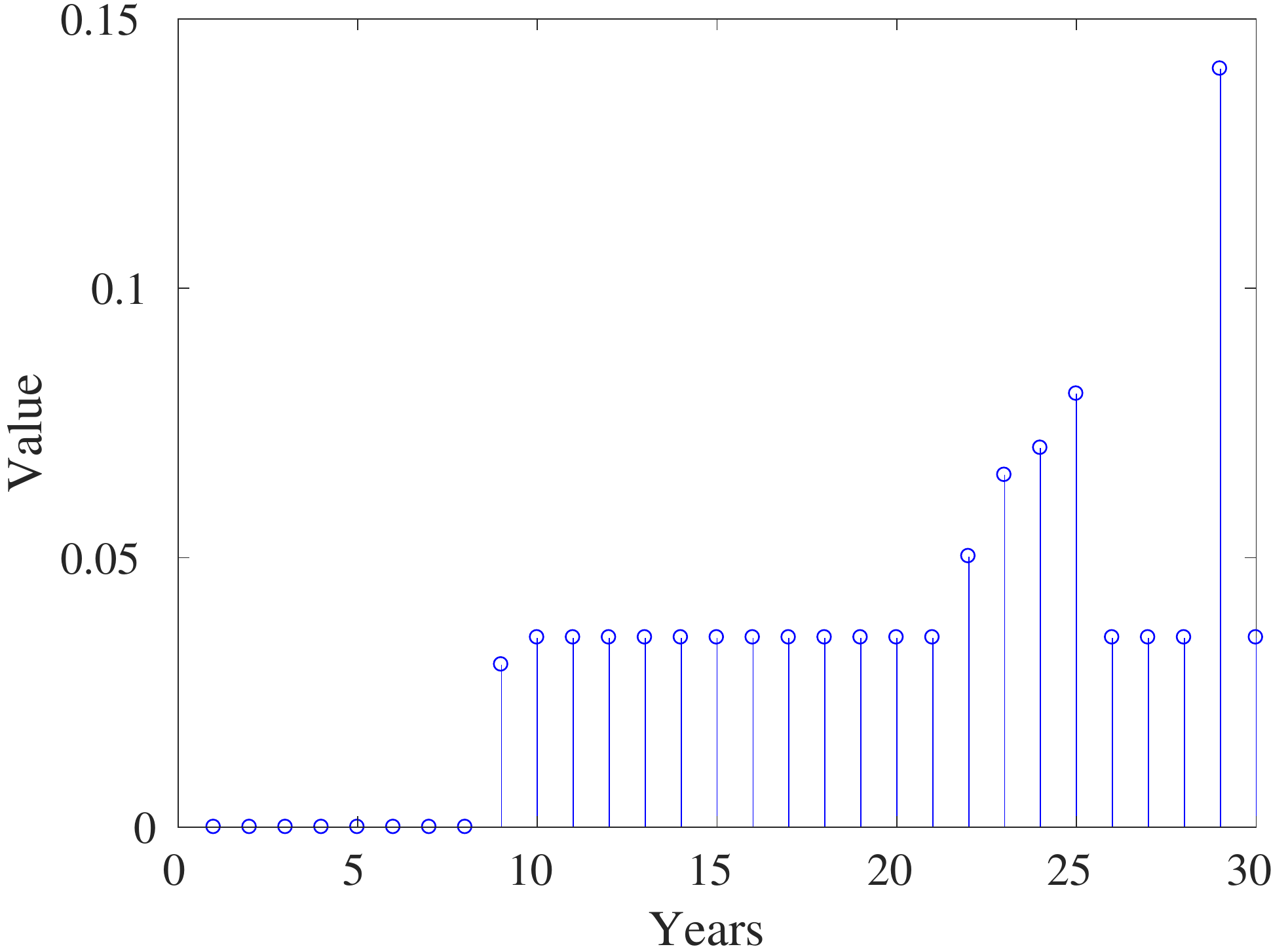}&
		\includegraphics[width=6cm]{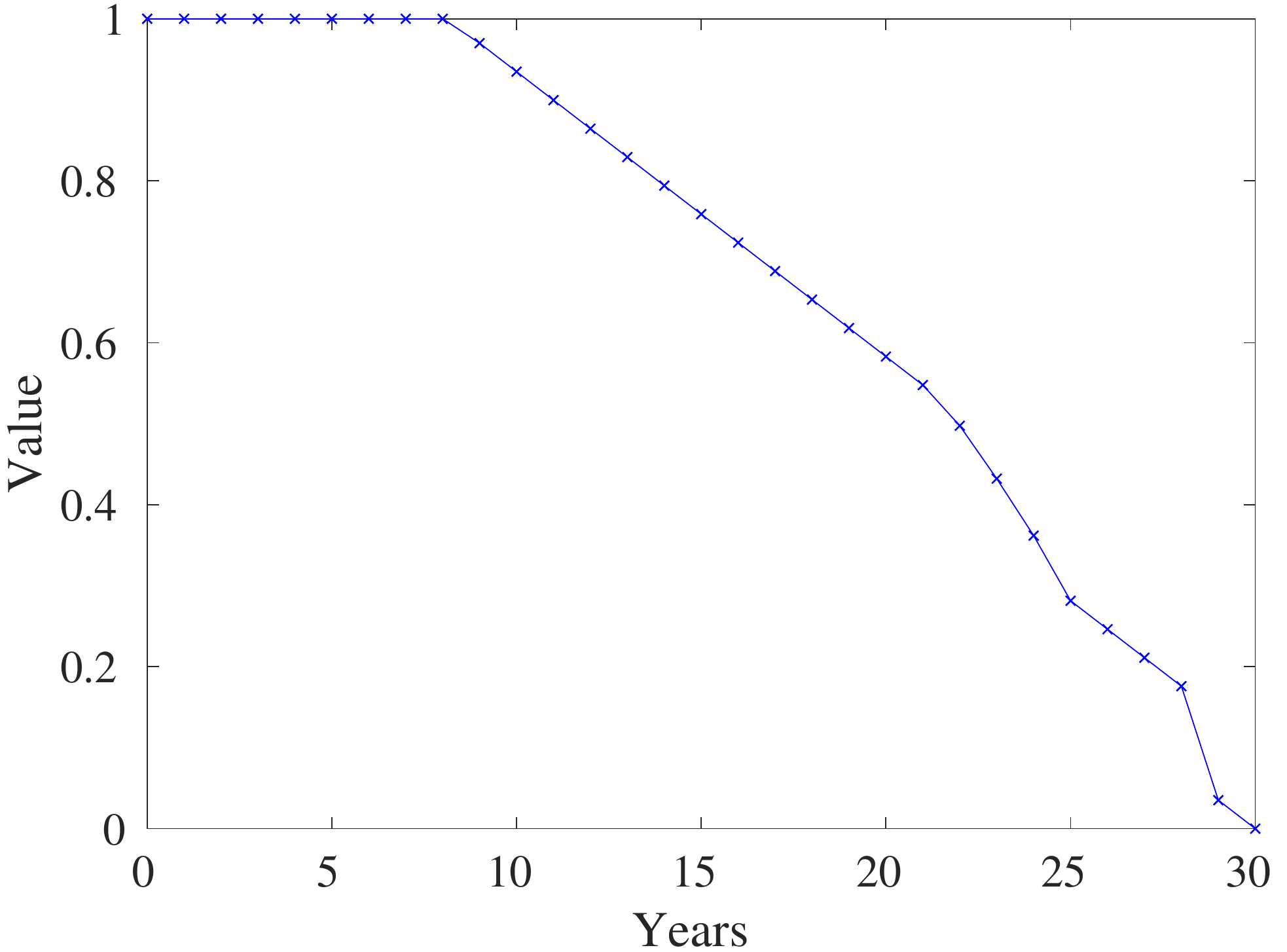}\\
		(c) withdrawals & 
		(d) nominal guarantee account \\
	\end{tabular}
	\caption{Value processes associated with the VA product when the pricing and hedging are performed based on the MMM and BA assuming an optimal withdrawal behavior, while the actual withdrawal behavior follows the BSM under risk-neutral pricing.}
	\label{bsmmfig}
\end{figure}

It is interesting to see what happens in a reversed scenario, where the VA provider prices and hedges under the MMM and BA, and the policyholder makes optimal withdrawals according to the BSM and risk-neutral pricing. Without repeating the detailed description of this scenario, the outcomes are shown in Figure~\ref{bsmmfig}, where the reserve account was managed recognizing the long-term growth under the MMM and BA, leading to a higher level of wealth accumulation than the nominal wealth account, and a surplus of 1.05 Million. The withdrawal behavior of the policyholder is similar to the first case considered, with a rather uniform withdrawal a few years into the contract term.

\begin{figure}[htp!]
	\centering
	\begin{tabular}{cc}
		\includegraphics[width=6cm]{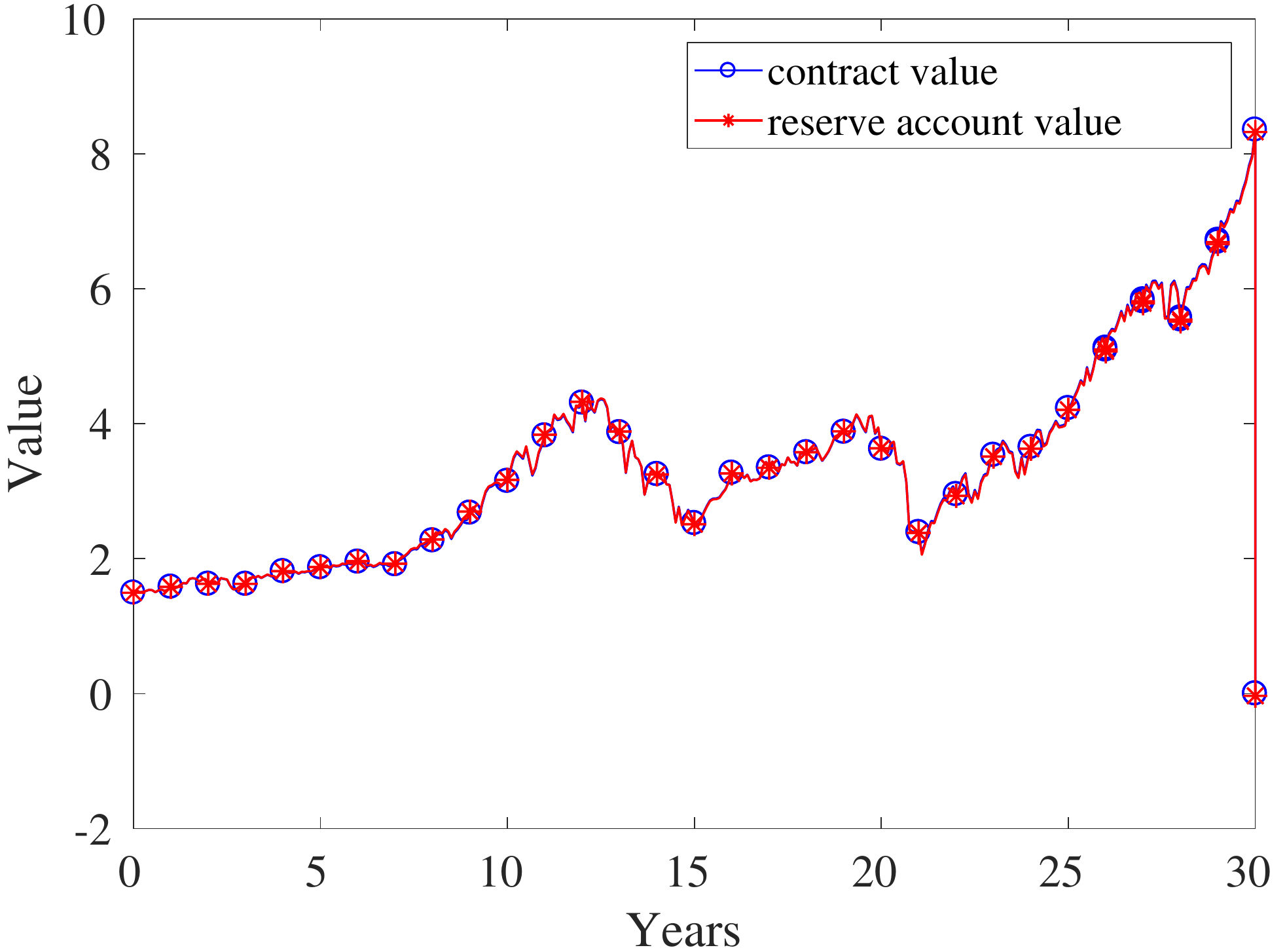}&
		\includegraphics[width=6cm]{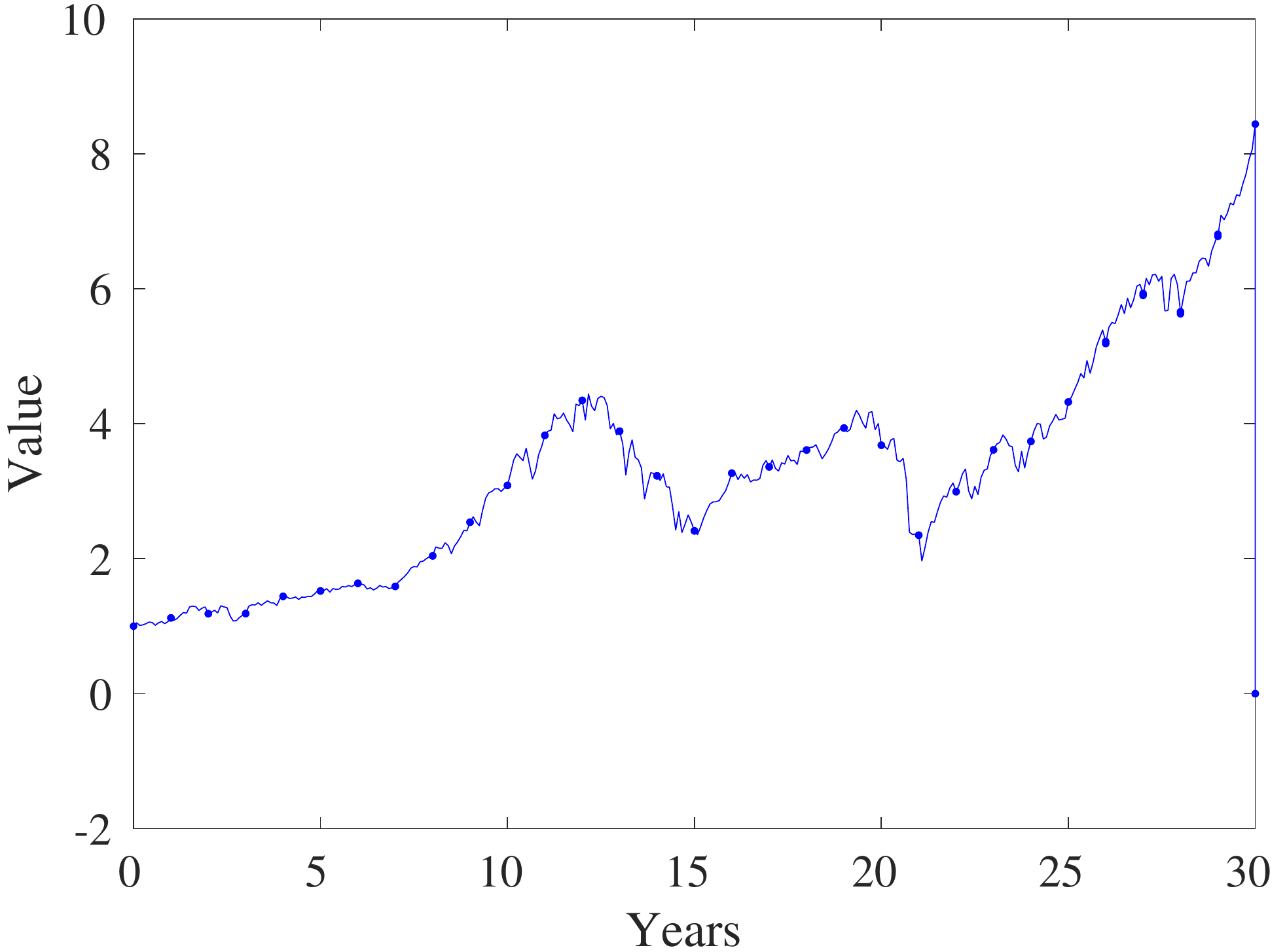}\\
		(a) contract value and reserve account  & 
		(b) nominal wealth account \\
		\includegraphics[width=6cm]{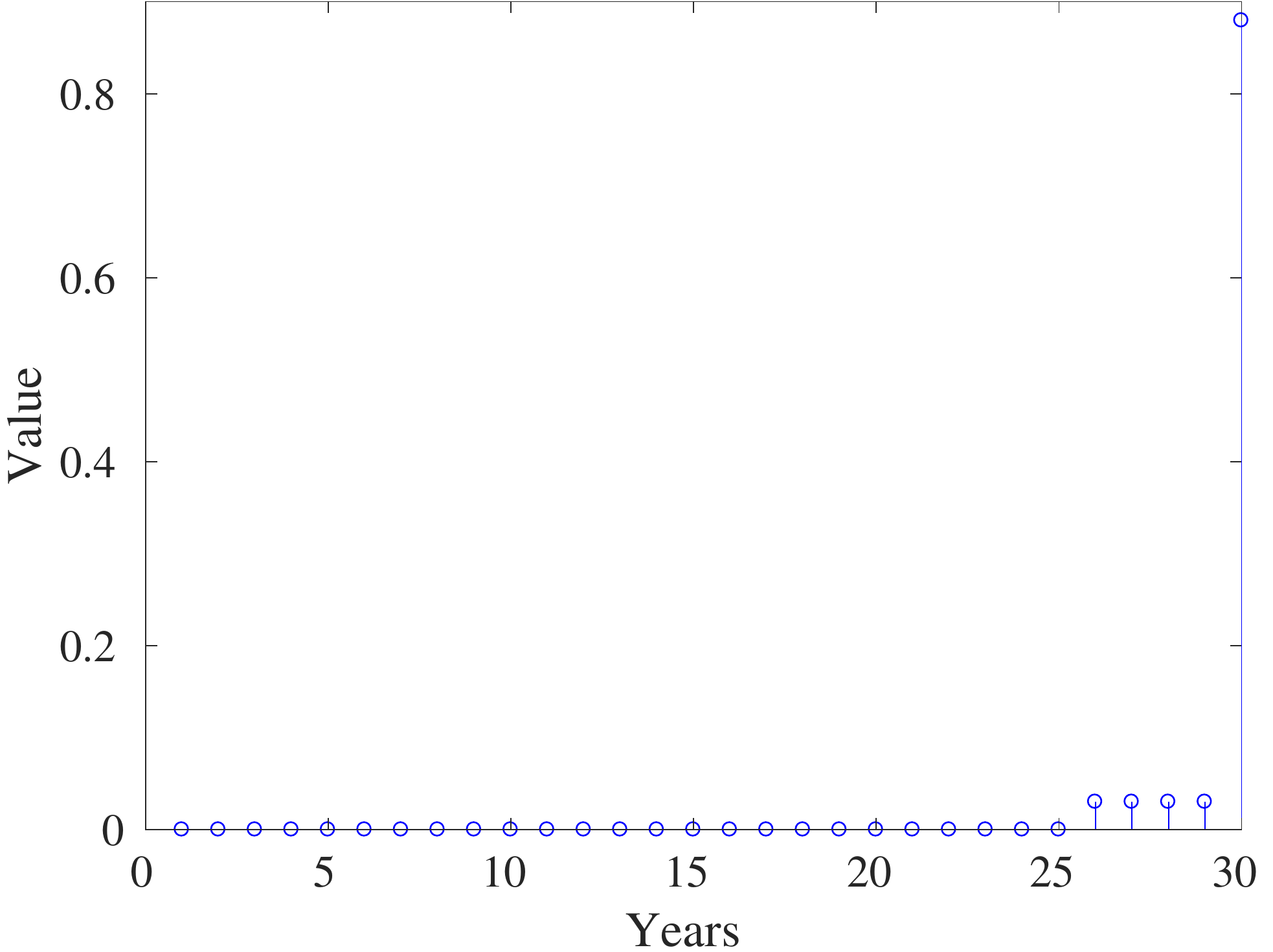}&
		\includegraphics[width=6cm]{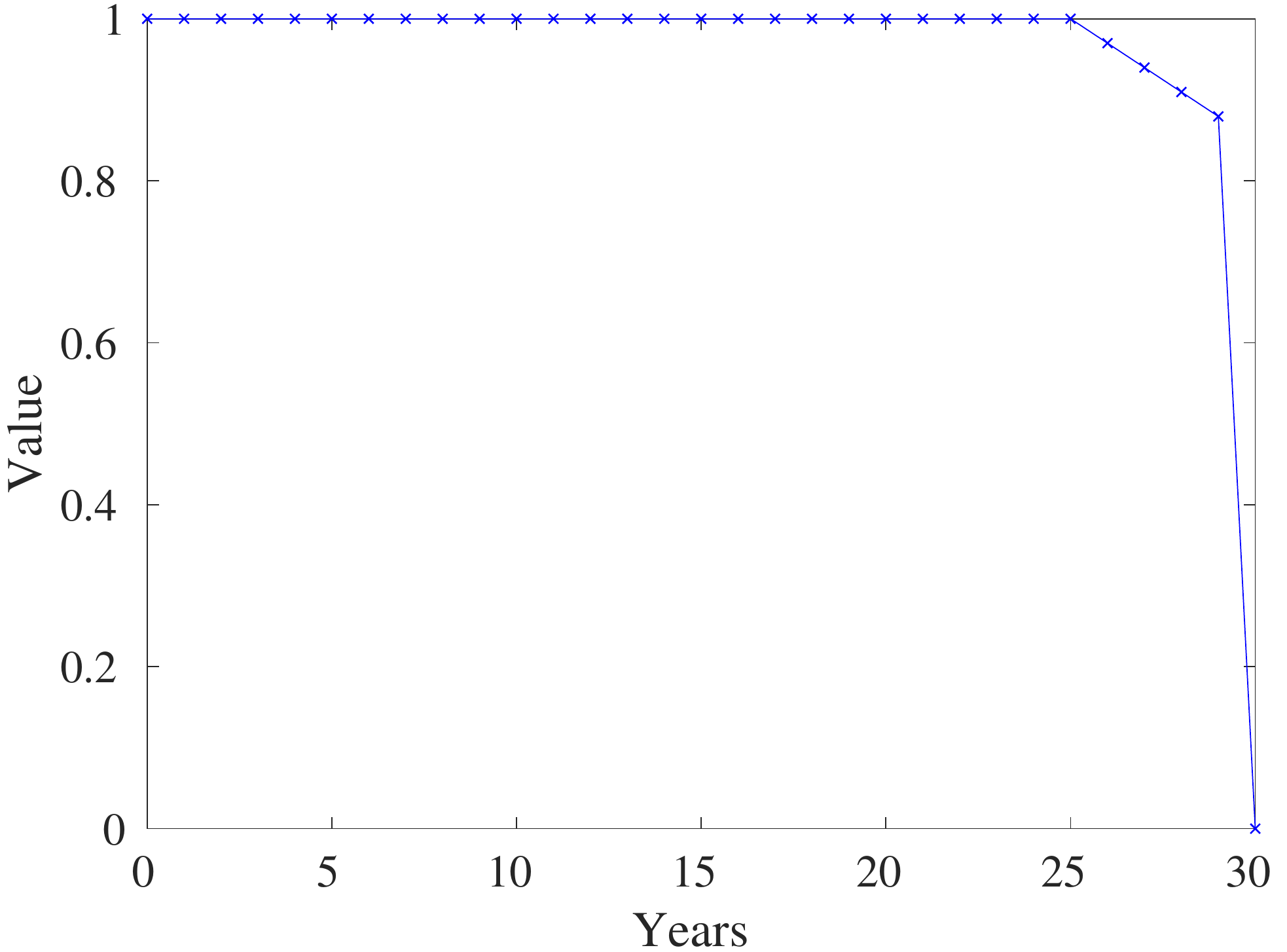}\\
		(c) withdrawals & 
		(d) nominal guarantee account \\
	\end{tabular}
	\caption{Value processes associated with the VA product when the pricing and hedging as well as the optimal withdrawals are performed based on the MMM and the BA.}
	\label{mmfig}
\end{figure}

Finally, to complete the empirical study, we consider the scenario when both the VA provider and the policyholder follow the MMM under the BA. The outcomes are shown in Figure~\ref{mmfig}. It can be seen that the VA provider in this case successfully hedged the VA product, ending up with a small deficit of 0.0319 Million in the reserve account. 

When comparing the two optimal withdrawal strategies based on the two pricing models, then one realizes that the strategy based on the BSM and risk-neutral pricing realizes a total withdrawal of 7.41 Million (of the locally risk-free security) over time, more than the total withdrawal of 6.29 Million following the static strategy. On the other hand, the optimal strategy based on the BA using the MMM realized a total withdrawal of 8.47 Million, which is producing more than 1 Million units of the savings account, thus, doubling the initial investment when counted in units of the savings account. This is considerably more than the risk-neutral approach delivered using the BSM. The reason is that the BSM under risk-neutral pricing creates a significant model error, which neglects the significantly positive long-term growth rate of the S\&P500 over that of the  savings account and, thus, gives major investment potential away. By changing the production method in the area of VAs as demonstrated, significantly higher returns on investment can be achieved.

\section{Conclusions}
\label{conc}
We considered the pricing of a variable annuity (VA) with GMWB under the benchmark approach (BA), where a classical equivalent risk-neutral pricing measure may not exist. We employed the real-world pricing formula to compute the value of the VA contract with GMWB under the minimal market model (MMM), and the associated hedging strategy for the VA product. We compared our results with those under the classical Black-Scholes model (BSM) under the risk-neutral pricing framework, through empirical backtests on the historical prices of the S\&P500 total return index, which were taken as the underlying of the VA product. 

From the empirical studies, we found that the VA provider can successfully hedge the VA product when the VA provider and the policyholder both employ the same pricing model to make hedging and withdrawal decisions. When the policyholder took a static withdrawal strategy without optimization, the VA provider ended up with a surplus. When the VA provider relied on the MMM and the BA to make hedging decisions and the policyholder took the BSM and the risk-neutral approach, the VA provider ended up with a surplus. However, when the VA provider took the BSM and risk-neutral approach, and the policyholder relied on the MMM and the BA, the VA provider ended up with a deficit. Our empirical studies show that the BSM and risk-neutral approach to the VA pricing problem may not be appropriate, in that when a sophisticated policyholder armed with a more accurate model such as the MMM, the VA provider risks having a significant deficit in the hedging of the VA product.

\bibliographystyle{elsart-harv}
\bibliography{allbibs}

\appendix
\section{A Brief Overview of the Benchmark Approach}
\label{ba}
In this section, we give a brief overview of the BA under a diffusive market model. Much of this section follows from \citet{PlatenHe06}, to which interested readers are referred  for a more complete treatment. 

We consider a general diffusive financial market model with uncertainties driven by a d-dimensional Brownian motion $\bW$, with $\bW(t)=\{(W(t)^1,...,W(t)^d)^\top,\,t\in[0,T]\}$, 
defined on a filtered probability space $(\Omega,\Fc,{\Ff},\Pmeas)$, where $T$ is some fixed time horizon, and the filtration $\Ff=\{\Fc_t,\,t\in[0,T]\}$ satisfies the usual conditions of right continuity and completeness, and models the accumulation of information over time; see \citet{KaratzasSh88}. We assume that there exists a locally risk-free savings account $S^0(t)$ and $m$ nonnegative risky primary security accounts $\bS(t)=(S^1(t),...,S^{m}(t))^\top$ satisfying the vector stochastic differential equation (SDE)
\beql{primary}
{d\bS(t)}={\bS(t)}\Prt{a(t)dt+b(t)\cdot d\bW(t)},
\quad t\in[0,T],
\eeq
where $a(t)$ is the instantaneous drift vector and $b(t)$ the instantaneous volatility matrix,  which both are assumed to be predictable and such that a unique strong solution of the above system of SDEs exists. We assume that all dividends and interests are reinvested. Without loss of generality, we further assume that $S^0(t)\equiv1$. This means that we denominate all security prices in units of  the locally risk-free savings account. In practice, the locally risk-free savings account may be approximated by the money market account that invests in short-term T-bills in a rolling manner. Thus in our notation, all primary security accounts are discounted by the locally risk-free savings account.

We denote by $S^{\bpi}$ the value process of a strictly positive, self-financing portfolio with portfolio weights $\bpi(t)=(\pi^1(t),...,\pi^m(t))^\top, t\in[0,T]$, which invests at time $t$ a fraction $\pi^j(t)$ of the total wealth in the $j$th primary security account, and the remaining wealth in the locally risk-free savings account. The value process satisfies then the SDE
\beql{port}
\frac{dS^{\bpi}(t)}{S^{\bpi}(t)}=\bpi(t)^\top(a(t)dt+b(t)\cdot d\bW(t)),
\quad t\in[0,T].
\eeq
By Ito's formula, the SDE for the log-price is of the form
\beql{lport}
d\log S^{\bpi}(t)=\bpi(t)^\top\Prt{\Prt{a(t)-\hf b(t)b(t)^\top\bpi(t)}dt+b(t)\cdot d\bW(t)},
\quad t\in[0,T].
\eeq
We consider the growth-optimal portfolio (GP) $S^{\bpi^*}$ of this investment universe for which the instantaneous expected growth rate, that is, the drift of (\ref{lport}), is maximized for all $t$. This is achieved by setting the optimal portfolio weights $\bpi^*(t)$ to 
\beql{pigp}
\bpi^*(t)=\underset{\bpi}{\arg\max}\,\bpi^\top\Prt{a(t)-\hf b(t)b(t)^\top\bpi},
\quad t\in[0,T].
\eeq
We assume that a solution to (\ref{pigp}) exists a.s. for all $t\in[0,T]$. One potential such solution is given by
\beql{optpi}
\bpi^*(t)=\Prt{b(t)b(t)^\top}^+a(t), \quad t\in[0,T], 
\eeq
where $\Prt{b(t)b(t)^\top}^+$ denotes the Moore-Penrose generalized inverse of the self-adjoint matrix $b(t)b(t)^\top$. Note that the value process of the GP is unique, however, the fractions may vary due to potential redundancies in the primary security accounts

For the market model to be viable, we assume that the GP process, denoted as $S(t):=S^{\bpi^*}(t), t\in[0,T]$, with $\bpi^*(t)$ given by (\ref{optpi}), exists and is strictly positive. By substituting (\ref{optpi}) into (\ref{port}), we obtain the SDE
\beql{gpp}
\frac{dS(t)}{S(t)}=\|\bth(t)\|^2dt+\bth(t)\cdot d\bW(t),
\quad t\in[0,T],
\eeq
where $\bth(t)=b(t)^\top\bpi^*(t)$. The above SDE can further be written as
\beql{gpm}
dS(t)=\alpha(t)dt+\sqrt{\alpha(t)S(t)}dB(t),
\quad t\in[0,T],
\eeq
where the drift $\alpha(t)=\|\bth(t)\|^2S(t)$ is assumed to be strictly positive, and $B^*(t)$, defined by the SDE
\beql{gpw}
dB(t)=\frac{\bth(t)}{\|\bth(t)\|}\cdot d\bW(t)
\quad t\in[0,T],
\eeq
with $B^*(0)=0$, forms a standard Brownian motion by Levy's characterization theorem.  So far, we only re-parametrized the GP dynamics different to the common volatility modeling specification. Note that the above drift $\alpha(t)$ can be, at this stage, still very general. Later on, we will make this drift more specific, which yields then a proper model.

The GP is the unique portfolio which, when used as numeraire or benchmark, makes any benchmarked portfolio process $\Sh^{\bpi}$, defined as $\Sh^{\bpi}(t)=\frac{S^{\bpi}(t)}{S(t)}$, a local martingale. If we assume the portfolio process to be nonnegative, the benchmarked portfolio process becomes a supermartingale by Fatou's lemma.
Given an $\Fc_T$-measurable nonnegative contingent claim $H(T)\ge0$ with maturity $T$, its, so called,  fair price process under the BA is given by the real-world pricing formula as
\beql{rwp}
H(t)=E_t\Prt{\frac{S(t)}{S(T)}H(T)},
\quad t\in[0,T],
\eeq
where $E_t(\cdot)=E(\cdot|\Fc_t)$ denotes the $\Fc_t$-conditional expectation under the real-world probability measure $\Pmeas$. 
The benchmarked fair price process, defined as $\hat{H}(t)=\frac{H(t)}{S(t)}$, forms then a nonnegative $(\mathbb{F},\Pmeas)$-martingale. The benchmarked fair price process $\hat{H}$, if replicable, represents the least expensive portfolio among all benchmarked nonnegative self-financing replication portfolios, which form supermartingales. 

\section{Modeling the Underlying Equity Index}
\label{models}
In this section we specify the model parameters of the SDE governing the underlying equity index (\ref{index}). The model parameters are described respectively under the risk-neutral pricing framework and the BA. As mentioned in Section \ref{va}, we take the locally risk-free security account as the numeraire, and consider all prices denominated in units of the locally risk-free savings account. 

\subsection{The Black-Scholes model}
The classical BSM is probably the most widely-known model to describe the price dynamics of a risky security within the framework of risk-neutral pricing theory. Under the BSM, the drift $\alpha(t)$ in the general formulation (\ref{gpm}) is modelled as $\alpha(t)=\sigma^2S(t)$, where $\sigma$ is the constant volatility parameter, and the equity index follows under the real-world probability measure $\Pmeas$ the SDE
\beql{Eqn:BS}
d S(t) = \sigma^2 S(t)\, dt + \sigma S(t)\, dB(t),
\eeq
where $B$ is a standard $\Pmeas$-Brownian motion. The equity index thus follows the geometric Brownian motion
\beql{bsm}
S(t)=S(0)\exp\Prt{{\hf\sigma^2} t+\sigma B(t)}, \quad {t\in[0,T]}.
\eeq

Following the standard procedures of Girsanov's theorem, the BSM admits a unique equivalent risk-neutral pricing measure $\Qmeas$, with the Radon-Nikodym derivative given by
\beql{bssdf}
Z(t)=e^{-{\sigma}B(t)-\frac{\sigma^2}{2}t}={\frac{S (0)}{S(t)}},\quad t\in[0,T],
\eeq
which is seen to be of the same form as the discount factor in (\ref{rwp}), as expected. Under the risk-neutral measure $\Qmeas$, the index $S$ is driftless and satisfies
\beql{bsmq}
dS(t)=\sigma S (t)dB^\Qmeas(t), \quad {t\in[0,T]},
\eeq
where $B^\Qmeas(t)=B(t)+{\sigma}t$ is a standard $\Qmeas$-Brownian motion.

\subsection{The minimal market model}
\label{formulation}
The minimal market model (MMM), see \citet{Platen01a}, is a stylized model under the BA. The MMM is incompatible with the risk-neutral pricing framework, in that an equivalent risk-neutral probability measure cannot exist.
When we apply the MMM, we make two important assumptions. First, the GP of a given investment universe is, generally, difficult to construct. However, it is shown by \citet{PlatenRe12an} that the GP of a stock market can be approximated by a respective well-diversified equity index. The MMM assumes that the equity index is a good proxy for the GP. Second, the drift coefficient $\alpha(t)$ in the general GP model (\ref{gpm}) is theoretically a complicated process depending on the instantaneous market prices of risks, and is, thus, difficult to specify. The MMM makes a critical simplification by assuming that $\alpha(t)$ is a simple deterministic exponential function $\alpha(t)=\alpha_0e^{\eta t}$. As a result, under the MMM, the GP $S(t)$ follows a time-transformed squared Bessel process of dimension four with a deterministic time transformation and satisfies the SDE
\beql{smm0}
dS(t)=\Prt{\alpha_0 e^{\eta t}}dt+\sqrt{\alpha_0e^{\eta t}{S(t)}}dB(t), \quad {t\in[0,T]}
\eeq
under the real-world probability measure $\Pmeas$; see \citet{RevuzYo99}. Here $\eta$ is the long-term expected growth rate of the equity index, and $\alpha_0$ is a constant representing the initial scale of the index. 

We define in the model the normalized GP as $Y(t)=\frac{\eta}{\alpha_0e^{\eta t}}S(t),t\in[0,T]$, which satisfies the SDE
\beql{smmy}
dY(t)=(1-Y(t))\eta dt+\sqrt{Y(t)\eta}dB(t), \quad {t\in[0,T]}.
\eeq
The normalized GP is seen to be mean-reverting around the level $1$. The mean-reversion of the normalized index implies a ``trend reversion" of the GP around its long-term exponential growth. It is well documented that a well-diversified index such as the S\&P500 index moves in the long-run in a trend reverting pattern, where the trend is usually interpreted as a slowly moving ``fundamental value" process; see \citet{shiller2015}. By decomposing the GP into the normalized GP index $Y(t)$ and a simple exponential fundamental value function $\frac{\alpha_0e^{\eta t}}{\eta}$, the MMM parsimoniously captures this important stylized fact. Furthermore, the instantaneous squared volatility of the normalized GP equals that of the GP and is inversely proportional to the value of the normalized GP, generating the so-called leverage effect.

The normalized GP described by (\ref{smmy}) is a time transformed square-root process of dimension four, with the transition law of a noncentral Chi-squared (NCX$^2$) distribution, given by
\beql{sb4}
Y(u)\eqbyl\frac{1-e^{-\eta(u-t)}}{4}
\chi^2_4\Prt{\frac{4 e^{-\eta(u-t)}}{1-e^{-\eta(u-t)}}Y(t)},
\quad 0\le t<u\le T,
\eeq
where $\chi^2_4(\zeta)$ denotes a NCX$^2$ random variable of degree $4$ and noncentrality parameter $\zeta$; see, e.g., \citet{BroadieKa06}. The $\chi^2_4(\zeta)$ random variable has finite moments of all positive orders. The probability density function (PDF) is given by
\beql{ncx2den}
f(\zeta,x) = 
\hf e^{-\frac{\zeta+x}{2}}
\Prt{\sqrt{\frac{x}{\zeta}}}
I_1\Prt{\sqrt{\zeta x}},
\quad x>0,
\eeq
where $I_1(\cdot)$ is the first order modified Bessel function of the first kind, see, e.g., \citet{RevuzYo99}. The transition density function of the normalized index is, thus, given by
\beql{ppd}
p\Prt{t,Y(t);u,Y(u)}=\frac{4}{1-e^{-\eta(u-t)}}f\Prt{\frac{4 e^{-\eta(u-t)}}{1-e^{-\eta(u-t)}}Y(t),\frac{4}{1-e^{-\eta(u-t)}}Y(u)}.
\eeq

It is worth mentioning that the normalized GP is dimensionless, and serves as a nontrivial state variable in that the transition density over the time period $(t,u)$ depends nonlinearly on the current state $Y(t)$. This is in marked contrast to the BSM, where the current price serves as a scaling factor of a time-homogeneous geometric Brownian motion. An implication of this dependence is that, unlike the BSM, (\ref{smm0}) is neither scale nor time invariant. Note however, the GP, as a time transformed squared Bessel process, has some self-similarity property.

\section{Pricing of the VA with Guarantee}
\label{pricing}
In this section we consider the pricing of the VA product described in Section \ref{va}. For the pricing of a given set of cash flows, we adopt the concept of a stochastic discount factor (SDF), where the present value of the cash flows are given by the sum of their expected values, after discounting by the SDF, conditional on all current information. Both risk-neutral pricing and pricing under the BA can be formulated in terms of an appropriate SDF. For more information on risk-neutral pricing theory, see \citet{delbaen06math} for an account. For a brief description of the BA, see \ref{ba}. 

To price the VA with guarantee value function $V(0,\bX(0))$, we note that no withdrawal is made between any withdrawal dates $(\pt,\tn)$, and that the wealth account is self-financing within this period. This leads to the following recurrence relation for the (left-continuous with right-hand limits) guarantee value function,
\beql{vrecur}
V(\pt^+,\bX(\pt^+))=E_{\pt^+}\Prt{D(\pt,\tn)V(\tn,\bX(\tn))},
\eeq
where $E_t(\cdot)=E(\cdot|\bX(t))$ is the expectation under the real-world probability measure $\Pmeas$, conditional on the current information represented by $\bX(t)$, and $D(t,u),0\le t<u\le T$ is the SDF over $(t,u)$. Under the BA, the SDF is given by $D(t,u)=\frac{S(t)}{S(u)}$, i.e., the ratio of the inverse GP. Under the risk-neutral pricing framework, the SDF $D(t,u)=\frac{Z(u)}{Z(t)}$, with $Z(t):=E_t\Prt{\frac{d\Qmeas}{d\Pmeas}}$ being the Radon-Nikodym derivative of the measure change from the real-world probability measure $\Pmeas$ to the equivalent risk-neutral measure $\Qmeas$, conditional on all available information at $t$. 

Upon withdrawal at time $\tn$, $0<n<N$, the left-hand limit of the value function satisfies the following jump condition
\beql{vjump}
V(\tn,\bX(\tn))=V(\tp,\bX(\tp))+C(\gamma_n,\bX(\tn)).
\eeq
In other words, the guarantee value immediately before the withdrawal is the sum of the value immediately after the withdrawal and the cash flow of the withdrawal. The active policyholder follows a dynamic strategy, which we obtain as the solution of the respective stochastic control problem. That is, for $0<n<N$, the withdrawal amount $\gamma_n$ is chosen according to the following total value maximizing strategy,
\beql{vmax}
\gamma_n=\Gamma(\tn,\bX(\tn))=
\underset{0\le\gamma\le A(\tn)}{\arg\max}\Brc{V(\tp,\bX(\tn)\setminus\gamma)+C(\gamma,\bX(\tn))},
\eeq
where $\bX(\tn) \setminus \gamma$ denotes the state variables $\bX(\tp)$ after withdrawal $\gamma$ is made, given the value of the state variables $\bX(\tn)$ before the withdrawal. On the other hand, the passive policyholder follows a static strategy of pre-determined withdrawal values. 
The contract value $V(0,\bX(0))$ can, thus, be computed recursively from (\ref{Vterm}), (\ref{ajump}), (\ref{wjump}) (\ref{vjump}) and  (\ref{vrecur}), along with the chosen strategy. These  procedures are summarized in Algorithm \ref{algo1}. 
\begin{algorithm}
	\caption{Recursive computation of $V(0,\bX(0))$}
	\begin{algorithmic}[1]
		\STATE initialize $V(T^+,\bX(T^+))=0$
		\STATE set $n=N$
		\WHILE{$n>0$}
		\STATE compute the withdrawal amount $\gamma_{n}$ with the optimal strategy (\ref{vmax}) or a pre-determined static strategy
		\STATE compute $V(\tn,\bX(\tn))$ by applying jump condition (\ref{vjump}) with appropriate cash flows
		\STATE compute $V(\pt^+,\bX(\pt^+))$ by computing (\ref{vrecur}) with terminal value $V(\tn,\bX(\tn))$ and the appropriate SDF $D(\pt,\tn)$
		\STATE set $n=n-1$
		\ENDWHILE
		\STATE return $V(0,\bX(0))=V(0^+,\bX(0^+))$
	\end{algorithmic}
	\label{algo1}
\end{algorithm}
To evaluate  (\ref{vrecur}), we discretize the underlying risk factors and approximate the conditional expectation in (\ref{vrecur}) using a finite sum. For the BSM, the risk factor is taken as the scaled Brownian motion $\frac{B(t)}{\sqrt{t}}$. For the MMM, the risk factor is the normalized GP $Y(t)$. Both risk factors have closed-form transition densities to facilitate the numerical computations.  
The contract value under the dynamic strategy is bounded from below by the corresponding value from any simple strategy such as the static one. If a closed-form pricing formula for the simple strategy value is available, Algorithm~\ref{algo1} can be easily modified to compute the optimal withdrawal premium on top of the suboptimal value of the simple strategy.

\end{document}